\documentclass[journal]{IEEEtran}
\usepackage[utf8]{inputenc}
\usepackage{hhline}
\usepackage{dsfont}
\usepackage{mleftright}
\usepackage{blindtext}
\usepackage{datetime}
\usepackage{amsmath,amssymb,amsfonts}
\usepackage{algorithmic}

\usepackage{textcomp}
\usepackage{xcolor}
\usepackage[ruled,vlined]{algorithm2e}
\usepackage{caption,subcaption}

\usepackage{mathtools}
\usepackage{graphics}
\def\BibTeX{{\rm B\kern-.05em{\sc i\kern-.025em b}\kern-.08em
    T\kern-.1667em\lower.7ex\hbox{E}\kern-.125emX}}
\usepackage{mathrsfs}
\usepackage{cite}
\usepackage{makecell}
\usepackage{commath}
\usepackage{multirow}  
\usepackage{graphicx}
\usepackage{comment}
\usepackage{diagbox}

\usepackage{accents}

\usepackage[utf8]{inputenc} 
\usepackage[T1]{fontenc}    
\usepackage{hyperref}       
\usepackage{url}            
\usepackage{booktabs}       
\usepackage{amsfonts}       
\usepackage{nicefrac}       
\usepackage{microtype}      
\usepackage{lipsum}
\definecolor{lightblue}{RGB}{153, 194, 255}

\graphicspath{ {./images/} }
\usepackage{tikz}%
\usetikzlibrary{arrows}%
\tikzstyle{startstop} = [rectangle, rounded corners, minimum width=3em, minimum height=2.5em, text centered, draw=black, fill=red!30]%
\tikzstyle{process} = [rectangle, minimum width=5em, minimum height=2.5em, text centered, draw=black]%
\tikzstyle{arrow} = [ultra thick,->,>=stealth]%
\tikzstyle{int}=[draw, fill=blue!15, minimum size=3.5em]%
\tikzstyle{init} = [pin edge={to-,thick,black}]%
\usepackage{amsthm}

\usepackage{bbm}

\def\x{{\mathbf x}}

\def\z{{\mathbf z}}

\def\R{{\mathbb R}}
\def\E{{\mathbb E}}

\DeclareMathOperator{\phase}{phase}
\DeclareMathOperator*{\argmin}{arg\,min}
\DeclareMathOperator*{\argmax}{arg\,max}
\usepackage{xcolor}

\usepackage{cite}
\usepackage{graphicx}

\usepackage{algorithmic}

\begin{document}
%
\title{A Sketching Framework for Reduced Data Transfer in Photon Counting Lidar}
%
%
%

\author{Michael P. Sheehan,~\IEEEmembership{Member,~IEEE},
         Juli\'an Tachella,~\IEEEmembership{Member,~IEEE}
        and~Mike E. Davies,~\IEEEmembership{Fellow,~IEEE}
\thanks{M.Sheehan, J.Tachella and M.Davies are with the School of Engineering, University of Edinburgh. This work is supported by  ERC Advanced grant 694888, C-SENSE and a Royal Society Wolfson Research Merit Award. Correspondence to M.Sheehan (Email: michael.sheehan@ed.ac.uk).}

}

%
%

\markboth{}%
{Shell \MakeLowercase{\textit{et al.}}: Bare Demo of IEEEtran.cls for IEEE Journals}
%



\maketitle

\begin{abstract}
Single-photon lidar has become a prominent tool for depth imaging in recent years. At the core of the technique, the depth of a target is measured by constructing a histogram of time delays between emitted light pulses and detected photon arrivals. A major data processing bottleneck arises on the device when either the number of photons per pixel is large or the resolution of the time-stamp is fine, as both the space requirement and the complexity of the image reconstruction algorithms scale with these parameters. We solve this limiting bottleneck of existing lidar techniques by sampling the characteristic function of the time of flight (ToF) model to build a compressive statistic, a so-called sketch of the time delay distribution, which is sufficient to infer the spatial distance and intensity of the object. The size of the sketch scales with the degrees of freedom of the ToF model (number of objects) and not, fundamentally, with the number of photons or the time-stamp resolution. Moreover, the sketch is  highly amenable for on-chip online processing. We show theoretically that the loss of information for compression is controlled and the mean squared error of the inference quickly converges towards the optimal Cram\'er-Rao bound (i.e. no loss of information) for modest sketch sizes. The proposed compressed single-photon lidar framework is tested and evaluated on real life datasets of complex scenes where it is shown that a compression rate of up-to 150 is achievable in practice without sacrificing the overall resolution of the reconstructed image.
\end{abstract}


%
\IEEEpeerreviewmaketitle

\section{Introduction}\label{Sec: Intro}
%
%
%
%

\begin{figure}[ht!]
\centering
\includegraphics[scale=0.25]{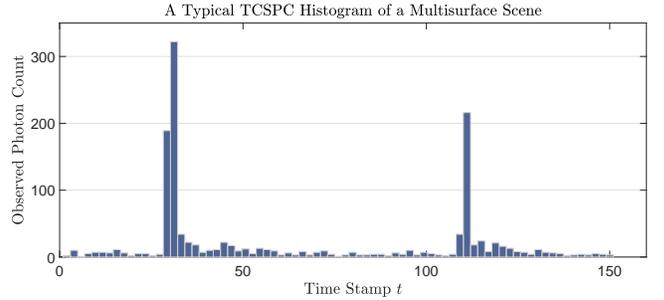}
\caption{An example of a TCSPC histogram of a pixel in a complex scene including a semi-transparent material (camouflage) in front of a person which is depicted by the 2 spikes of the histogram respectively.}
\label{fig: intro hist}
\end{figure}

\IEEEPARstart{S}{ingle} photon light detection and ranging (lidar) has emerged as an important depth imaging technique prevalent in the automobile \cite{Hecht:18,lidarauto}, defence \cite{6159363} and forestry industries\cite{PIERZCHALA2018217}. This modality has the unique advantage of offering very high depth resolution \cite{manipop} even at long-range scenes using low-power (eye-safe) lasers \cite{Pawlikowska2017SinglephotonTI}.  The technique has at its core the ability of emitting light pulses and detecting each single-photon as it arrives, thereby obtaining a depth estimate by measuring the round-trip time of individual photons. By using a time correlated single-photon counting (TCSPC) system, a histogram can be formed indicating the time delay between emitted light pulses and detected photons for each pixel, with a proportion of the photons originating from background or ambient light (e.g. the sun). The number of counts per time histogram bin provide information on the depth and reflectivity of the object or scene. The presence of a peak in the histogram indicates an object is present within the range of the lidar system. The location of this object corresponds to the location of the impulse response. If the material is semi-transparent (e.g. glass, water, camouflage) or the laser footprint is large, then multiple peaks with different intensities may exist within a single pixel \cite{manipop}. A standard example of a TCSPC histogram for a given pixel within a scene is shown in Figure \ref{fig: intro hist}.

\par The image restoration task reduces down to inferring the positions and intensities of the peaks in the histogram for each pixel in the image.  Typically, the time-correlated single-photon counting data is collected in two main approaches, either: (i) the time-stamp of each photon is recorded\cite{manipop}, or (ii) a temporal histogram, as seen in Figure \ref{fig: intro hist}, is constructed which counts the number of photons detected per histogram bin of time-interval $\Delta \tau$ \cite{30frames,3Dstacked}. In either case, the time-correlated single-photon counting data has to be recorded, stored in memory and transferred from the chip for each pixel in the scene.   The development of high rate, high resolution, low power ToF image sensors is challenging due to the large data volumes required. This causes a major data processing bottleneck on the device when either the
number of photons per pixel $n$ is large, the time resolution, $\Delta \tau$, is fine or the spatial resolution is high, as the space requirement, power consumption and computational burden of the depth reconstruction algorithms scale with these parameters \cite{tachellaNComms}.

\par Various existing methods have attempted to tackle the trade-off between depth resolution and computational/space complexities. A number of papers \cite{hardware,slidingGate,128x96,128x128,128x128b} propose methods to address the trade-off between depth resolution and the complexities associated with the TCSPC histogram. Henderson et al. \cite{hardware} propose a method that employs a gated procedure to coarsely bin the detected photons, whilst Ren et al. \cite{slidingGate} develop a sliding window approach to achieve high resolution depth. Walker et al. \cite{128x96} calculate the depth directly from the photon time-stamps. However in all of these approaches, the approximations formed on-chip compromise the depth resolution of the image. Della Rocca et al. \cite{128x128,128x128b} proposes to only collect the histograms of photon detections when there is a significant change of activity. This method reduces the data-transfer, as it is only required during specific moments in time. Similarly, Hutchings et al. \cite{3Dstacked} propose a method of discarding photon detections based on activity. However, these methods can potentially remain idle when there is a small change in activity, and can also suffer from a loss of temporal resolution due to coarse histogram binning. Zhang et al. \cite{30frames} propose a method of reducing the transfer of photon detections by performing a coarse to fine approximation of the ToF data. At each scale, a coarse histogram is constructed with a limited number of bins. Multiple histograms of increasing resolution have to be formed, hence the method has an increased  total acquisition time and can also suffer from a loss of temporal resolution.  In \cite{rapp2020dithered}, Rapp et al. proposed a subtractive dithering for SPAD arrays that increases depth resolution without increasing the overall time-stamp resolution. 
\par Compressive sensing strategies have been successfully applied to lidar \cite{7178153,halimi:hal-02298998}, focusing on compressing the information across pixels. Kadambi et al. \cite{7178153} propose to exploit the sparsity of natural scenes in some representation domain (e.g. wavelet transform) to reduce signal acquisition. The depth accuracy is limited by the level of amplitude noise and decay of the impulse response and is therefore limited to the case of one surface per pixel. Furthermore, the proposed method still requires large amounts of single-photon counting data to be transferred off-chip and therefore does not tackle the inherent data transfer bottleneck that we address in this paper. In a similar vein, Halimi et al. \cite{halimi:hal-02298998} propose an adaptive sampling strategy that is scene dependent. By building up regions of interest and data driven depth maps in an iterative manner, they efficiently choose suitable scan positions to reduce acquisition time by up to 8 times in certain scenarios. However, the method relies on building TCSPC histograms and solving a maximization problem at each iteration of their adaptive algorithm. The method therefore has limitations for real-time processing especially when the amount of single-photon counting data is large. These compressive sensing based methods perform compression within the spatial domain and not, in the case of our method, throughout the depth domain and are therefore fundamentally different in practice and can still suffer from data-transfer bottlenecks. The sketched-based method proposed in this paper is complimentary to the compressive sensing based approaches as one can compress along both the spatial and temporal simultaneously. Another approach to reduce the data transfer of the information needed to reconstruct the lidar image is to compress the data on-chip. As highlighted in \cite{maksymova2018review}, standard low-level data compression methods can be used to compress the data on-chip, however these methods can only offer up to a modest $50\%$ data reduction and in some cases involve significant on-chip computation or there are limitations with respect to on-chip storage.

\par In this paper, we propose a novel solution to this bottleneck of existing lidar techniques by calculating on-the-fly summary statistics of the photon time-stamps, a so-called
sketch, based on samples of the characteristic function of the  ToF model. Distinct to compressive sensing, the goal here is not to recover the photon counting data but rather the underlying probability distribution. In this sense, we are estimating the probability model directly from some summary statistics and therefore our proposed framework utilises much of the theory found in the generalised method of moments \cite{hansenGeMM,gemmhall}, empirical characteristic function \cite{10.2307/2958763,10.2307/2985144} and compressive learning \cite{gribonval2020statistical,keriven2018sketching,SheehanCICA} literature. The size of the sketch scales with the degrees of freedom of the ToF model (i.e., number of
objects in depth) and not with the number of photons or the fineness of the time resolution, without sacrificing precision in depth. The sketch can be computed for each incoming
photon in an online fashion, only requiring a minimal amount of additional computation which can be performed efficiently on-chip. The sketch can be shown to capture all the salient
information of the histogram, including the ability to explicitly remove background light or dark count effects, in a compact and data-efficient form, suitable for both on-chip processing or off-chip post processing. Furthermore, we develop a compressive lidar image reconstruction algorithm which has computational complexity dependent only on the size of the sketch. Our proposed
method paves the way for high accuracy 3D imaging at fast frame rates with low power consumption. In summary the main contributions of the paper are as follows:
    \begin{itemize}
        \item We propose a principled approach for compressing time-of-flight information in an online fashion without the requirement to form a histogram and without compromising depth resolution.
        \item A compressive single-photon lidar  algorithm is proposed which does not scale with either the number of photons or the time-stamp resolution in terms of space and time complexity.
        \item The statistical efficiency, given a compression rate (or sketch size), is quantified for different single-photon lidar scenarios, showing that only limited measurements of the characteristic function are needed to achieve negligible information loss. 
    \end{itemize}

The remainder of the work is organized as follows. Section 2 details the ToF lidar acquisition systems and the ToF observation model used in single-photon lidar and also presents the idea of summary statistics used for parameter estimation. In Section 3 we detail the construction of the sketch using two different sampling schemes and we further demonstrate how our sketched lidar approach can be implemented in an online processing manner. In Section 4 we detail our proposed compressive single-photon reconstruction algorithm that has computational complexity which scales with the sketch size $m$ as well as quantifying the statistical efficiency of the estimated parameters $\theta$. Results of the compressive lidar framework are analysed on both synthetic and real datasets in Section 5. Section 6 finally summarizes our conclusions and discusses future work.

\section{Background}\label{Sec: Background}
\subsection{Photon Counting Lidar Acquisition}

\begin{figure}[ht!]
\centering
\includegraphics[scale=0.2]{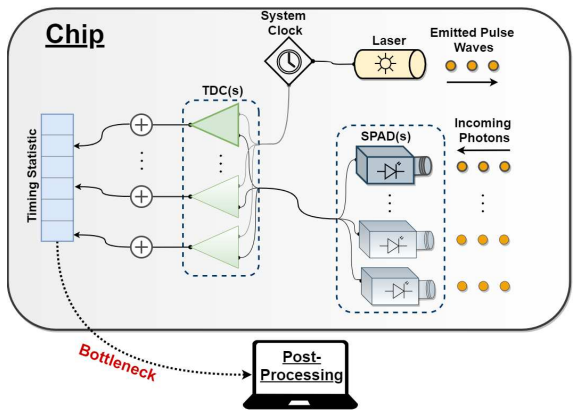}
\caption{A schematic of a typical lidar pixel where either one or multiple SPADs and TDCs are used.}
\label{Fig: Lidar Device}
\end{figure}

Figure \ref{Fig: Lidar Device} depicts a simplified schematic of a typical lidar device. A laser emits a pulse wave of photons to a scene that triggers the system clock where a single photon avalanche diode (SPAD) is then used to detect individual photons. The SPAD consists of a reverse-biased photodiode which, in the presence of a single photon, induces an avalanche of electrical charge carriers that are directly detectable. A time-to-digital converter (TDC) then converts the signal to a digital time-stamp that updates a timing statistic in an online manner (specific details of the timing statistic are discussed later in the section).\par Conventional lidar devices typically consist of a single SPAD and a pair of scanning mirrors that raster-scan points in a scene. In general, these approaches only register the first photon in the frame therefore multiple laser cycles are required to build an accurate timing statistic before traversing to a new point in the scene. In high ambient conditions, a significant pile-up effect can occur due to the dead-time of the single SPAD \cite{Gyongy:20}. This problem can be alleviated by using multiple SPADs in parallel resulting in multi-event time-stamp collection \cite{Gyongy:20}. In contrast to the first photon approach \cite{Krstajic:15}, SPADs and TDCs acting in parallel can register multiple photons per frame leading to a richer and larger timing statistic. Flood light illumination can also be used instead of slower raster-scan processes. As shown in Figure \ref{Fig: Lidar Device}, multiple SPADs connect to multiple TDCs to allow efficient parallelization, where each TDC accumulates digital time-stamps to the timing statistic.
\par The timing statistic terminology is used to refer to the various methods of collecting and storing the time-stamps that are then transferred off-chip to construct a depth map. The most common timing statistic used is a histogram, seen in Figure \ref{fig: intro hist}, that clusters the digital time-stamps into discretized bins. This method is most commonly used when the number of detected photons is large and can be devised as a compression in itself. In recent years, however, modern lidar devices can produce finer depth resolution causing a data transfer bottleneck as the histogram can become too large to transfer off-chip. To compensate, coarser bin widths can be used to reduce the size of the histogram \cite{hardware}, creating a trade-off between depth resolution and quicker data transfer. In contrast, if the number of photons detected is small, for example in the photon-starved regime \cite{manipop, Kirmani58}, it is more efficient, from a data processing point of view, to store the specific time-stamp of all the detected photons. In general, depth estimation is carried out off-chip as part of a post-processing stage. In this paper, we propose a novel online timing statistic that circumvents the need to either construct and store a large histogram or collect each individual photon time-stamp, leading to substantial compression. As the sketch is constructed only at the timing statistic stage in Figure \ref{Fig: Lidar Device}, any existing techniques that reduce pile-up, e.g. via (parallel) multi-event detection, can be readily used in conjunction with our proposed technique.

\subsection{Lidar Observation Model}\label{Subsec: Lidar Observation Model}
 Throughout this section, both the lidar observation model and the constructed sketches are discussed in terms of a single arbitrary pixel in the scene. Let $\tau$ denote the physical time-stamp such that the discretized time-stamp is denoted $t=\frac{\tau}{\Delta\tau}$.
Then, for an arbitrary pixel, the photon count at discretized time-stamp $t=0,1,\dots,T-1$ can be modelled as a Poisson distribution \cite{poissonmodel,altmann1}:
\begin{equation}
    \label{Eqn: Poisson Observ Model}
    y_{t_k}|(r,b,t_k) \sim \mathcal{P}(r h(t-t_k)+b),
\end{equation}
where $r\geq 0$ denotes the reflectivity of the detected surface, $h(\cdot)$ is the impulse response of the system, $b$ defines the level of background photons  and $t_k$ denotes the location of the $k$th surface in the pixel. The number of discretized time-stamp bins over the range of interest is denoted by $T$. For simplicity, here we assume that the integral of the impulse response $H=\sum_{t=0}^{T-1} h(t)$ is constant although the proposed approach can accommodate more complex scenarios. If the lidar system is in free running mode where multiple acquisitions of a surface/object are obtained, then the interval $[0,1,\dots,T-1]$  can be thought of as circular in the sense that time-stamp $T$ is equivalent to the time-stamp $0$. 
\par Alternatively, one can instead model the time of arrival of the $p$th photon detected  for a single pixel in the scene.  We assume there are $K$ distinct reflecting surfaces  within the pixel,  where $\alpha_k$ and $\alpha_0$ denote the probability that the detected photon originated from the $k$th surface and background sources, respectively.  Furthermore, it is assumed that for a single pixel, a total of $n$ photons are detected during the whole acquisition window of the lidar device . Let  $x_p=0,1,\dots,T-1$  denote the time-stamp of the $p$th photon where $1\leq p\leq n$, then $x_p$ can be described by a mixture distribution \cite{altmann2}
\begin{equation}
    \label{Eqn: Alternative mixture Observ model}
    \pi(x_p|\alpha_0,...,\alpha_{K},t_1,...,t_K)= \sum^K_{k=1}\alpha_k\pi_s(x_p|t_k)+\alpha_0\pi_b(x_p),
\end{equation} 
where $\sum^K_{k=0}\alpha_k=1$  and the symbol $\pi$ denotes a probability distribution over $x_p$ . The distribution of the photons originating from the signal and background  are defined by the distribution $\pi_s(x_p|t)=h(x_p-t)/H$ and the uniform distribution $\pi_b(x_p)=1/T$ over $[0,1,\dots,T-1]$ , respectively. Often in practice, the signal distribution $\pi_s$ is  modelled either using a discretized Gaussian distribution over the interval $[0,1,\dots,T-1]$  or through the  data driven impulse function which is calculated through experiments. In Section \ref{Sec: Experiments}, we consider both.
\subsection{Summary Statistics}\label{Subsec: Summary Statistics}
Our acquisition goal is to obtain parameter estimates  $\theta\coloneqq(\alpha_0,\alpha_1,\dots,\alpha_K,t_1,\dots,t_K)$  of the signal model in (\ref{Eqn: Alternative mixture Observ model}), given the time-stamp of photons detected  for each pixel in the scene.  Parameter estimation involves the inference of a set of parameters $\theta\in\Theta\subset\R^{2K+1}$ associated to a probability model $\pi(\cdot\mid\theta)$ defined on some space $\mathbf{x}\in\R^d$. In the case of single-photon counting lidar, the dimension $d=1$. Typically, we observe a finite dataset $\mathcal{X}=\{\mathbf{x}_i\}_{i=1}^n$ of $n$ samples which we assume is sampled from the distribution given in (\ref{Eqn: Alternative mixture Observ model}). Maximum likelihood estimation (MLE) is a traditional parameter estimation method whereby a likelihood function associated with the finite data is maximized with respect to the model parameters, e.g.
\begin{equation}
\label{Eqn : MLE}
    \hat{\theta} = \argmax_{\theta}\frac{1}{n}\sum^n_{i=1}\log \pi(\mathbf{x}_i\mid\theta).
\end{equation}
\subsubsection{Generalised Method of Moments}
In some cases, the likelihood function might not have a closed form solution nor a computationally tractable approximation \cite{hansenGeMM}. Generalised method of moments \cite{hansenGeMM,gemmhall} (GeMM) is an alternative parameter estimation method where one estimates $\theta$ by matching a collection of generalised moments with an empirical counterpart computed over a set of finite data sampled from the distribution $\pi(\x\mid \theta)$. Given a nonlinear function $g:\R^d\rightarrow\mathbb{C}^m$, then we define the expectation constraint 
\begin{equation}
\label{Eqn: GeMM constraint}
    \E g(\x;\theta)=0,
\end{equation}
where $\E$ denotes the expectation with respect to the probability distribution $\pi(\x \mid \theta)$. Typically, the GeMM estimator is obtained by minimising a quadratic cost of the empirical discrepancy with respect to $\theta$ to try impose the moment constraints of (\ref{Eqn: GeMM constraint}). Let us define
\begin{equation}
\label{eqn: empirical GeMM}
    g_n(\mathcal{X};\theta)\coloneqq \frac{1}{n}\sum^{n}_{i=1}g(\x_i;\theta),
\end{equation}
calculated over $\mathcal{X}=\{\x_i\}^n_{i=1}$, then a GeMM classically takes the form \cite{hansenGeMM,gemmhall}
\begin{equation}
    \label{Eqn: GeMM loss}
    \hat{\theta}\coloneqq \argmin_\theta  g_n(\mathcal{X};\theta)^T \mathbf{W}  g_n(\mathcal{X};\theta),
\end{equation}
where $\mathbf{W}$ is a symmetric positive definite weighting matrix that may depend on $\theta$. 

\subsubsection{Compressive Learning}
Building on the concept of GeMM, compressive learning \cite{gribonval2020statistical,keriven2018sketching} utilises generalised moments of the data but with the distinct goal of reducing signal acquisition, space and time complexities. The link to GeMM is established by separating the function $g$ into the following particular form:
\begin{equation}
    \label{Eqn : CL sep form GeMM}
    g(\x;\theta) = \Phi(\x)-\E_\theta\Phi(\x),
\end{equation}
where $\Phi:\R^d\mapsto\mathbb{C}^m$ is often referred to as the feature function. The separable form decouples the measured moments, $\Phi(\x)$, from the parameters $\theta$ that are to be estimated. This is not a usual assumption in GeMM, although it may arise in particular cases. By denoting the empirical mean or the so-called sketch as 
\begin{equation}
    \label{Eqn: The sketch}
    \z_n\coloneqq\frac{1}{n}\sum^{n}_{i=1}\Phi(\x_i),
\end{equation}
we can estimate $\theta$ solely from the sketch $\z_n$ by minimising
\begin{equation}
    \label{Eqn: CL loss function}
    \hat{\theta} = \argmin_\theta \lVert\z_n-\E_\theta\Phi(\x)\rVert^2_\mathbf{W},
\end{equation}
which is the particular compressive GeMM loss of (\ref{Eqn: GeMM loss}). In Section \ref{Sec: Sketched Lidar Reconstruction}, we explicitly define the weighting matrix $\mathbf{W}$ for compressive single-photon counting lidar. The separable form of $g$ in (\ref{Eqn : CL sep form GeMM}) allows a sketch statistic $\z_n$ to be formed with a single pass of the data without the need to store $\mathcal{X}$, and it can easily be updated on the fly with minimal computational cost. The sketch statistic has size $m$, or size $2m$ if decoupled into its real and imaginary components, which, fundamentally, scales independent of the dimensions of the dataset $\mathcal{X}$, which in the case of single-photon lidar is the photon count $n$ or the binning resolution $T$.
\subsubsection{Empirical Characteristic Function}
A specific type of GeMM is the empirical characteristic function (ECF) estimation\cite{10.2307/2958763,10.2307/2985144,carrasco2000generalization}, and occurs when the generalized moment is chosen to be $\Phi(\x)=[e^{ {\rm i}\omega_j^T\x}]_{j=1}^m$, where ${\rm i}=\sqrt{-1}$ and $\omega_j$ is a discrete set of frequencies. It is of particular interest as the expectation of $\Phi$, namely $\Psi_\pi(\omega)=\mathbb{E}_\theta e^{{\rm i}\omega^T\x}$, is specifically the characteristic function (CF) of the probability distribution $\pi(\x\mid \theta)$ at frequency $\omega$. The CF exists for all distributions, and often has a closed form expression. Moreover, it captures all the information of the probability distribution \cite{osomeFouriermethods}, therefore giving a one-to-one correspondence between the CF and the probability distribution $\pi(\x\mid\theta)$.  The CF also has the favourable property that it decays in frequency, i.e. $\Psi_\pi(\omega)\rightarrow 0$ as $\omega \rightarrow \infty$, under mild conditions on the probability distribution $\pi(\x\mid\theta)$ \cite{osomeFouriermethods,lukacs1952analytic}. For a single depth observation model in (\ref{Eqn: Alternative mixture Observ model}) (i.e. $K=1$) and a discrete impulse response function $h$, we define the characteristic function of the observation model 
\begin{align}
\begin{split}
    \label{Eqn: Char function Obs Model}
    \Psi_\pi(\omega)\;= \;& \alpha_1\Psi_{\pi_s}(\omega)+\alpha_0\Psi_{\pi_b}(\omega)\\
   \; =\; & \alpha_1\hat{h}(\omega)e^{{\rm i}\omega t}+\alpha_0 D_{\frac{T-1}{2}}(\omega)
    \end{split}
\end{align}

\noindent where $D_n(x)=\frac{\sin((n+1/2)x}{2\pi\sin(x/2)}$ is the Dirichlet kernel function \cite{Dirich_kernel} and $\hat{h}$ denotes the (discrete) Fourier transform of the impulse response function $h$. It should be noted that we could consider different distributions $\pi_b$, and hence CFs, to model the detected photons originating from more complex background sources,  for example highly scattering environments like fog. However, this is beyond the scope of this paper.  \par The feature function $\Phi$ is a complex valued function of size $m$. With regards to hardware implementation, it is often preferable and convenient to work directly with real valued functions. The complex term $e^{{\rm i}\omega x}$ can be alternatively written as $\cos(\omega x)+{\rm i}\sin(\omega x)$, where $e^{{\rm i}\omega x}$  has been decoupled into its real and imaginary components. As a result, the feature function $\Phi$ can be equivalently written as a real valued feature function  $\Phi:\R^d\mapsto\R^{2m}$, consisting of $2m$ real valued terms by stacking the real and complex components, for e.g. 
\begin{equation*}
\Phi(x)= \begin{bmatrix}\cos\left(\omega_1 x\right)\\\vdots\\\cos\left(\omega_m x\right)\\[0.5em]
    \sin\left(\omega_1 x\right)\\\vdots\\ \sin\left(\omega_m x\right)
\end{bmatrix}.  
\end{equation*}
For sake of fair comparison to existing hardware implemented methods in the literature, the results and figures presented represent a sketch of size $2m$, consisting of $2m$ real valued measurements. The nature of the feature function, in terms of it being represented by a complex or real valued function, will be made clear in its context throughout the paper.

\section{Sketched Lidar}\label{Sec: Sketched Lidar}
We start with a warm up example to highlight the potential of using a sketch for single-photon lidar and to motivate the design of the sketch sampling procedure which will be discussed in Section \ref{subsec sampling schemes}.
\subsection{Compressing Single Depth Data }\label{subsec: warm up}
In the absence of photons originating from background sources and the presence of a single surface or object, the sample mean of all the photon time-stamps ($\Phi(x)=x$) is the simplest summary statistic for estimating the single location parameter $t_1$. This only holds in the noiseless case as the sample mean estimate is heavily biased toward the centre of the histogram when background photons are detected. 
\par Suppose, we instead observe the cosine and sine of each photon count $x$ with angular frequency $\omega=\frac{2\pi}{T}$, namely
\begin{equation}
\label{eqn: motivate Phi}
    \Phi(x)= \begin{bmatrix}\cos\left(\frac{2\pi x}{T}\right)\\[0.5em]
    \sin\left(\frac{2\pi x}{T}\right)
\end{bmatrix},
\end{equation}
and denote $\z_n$ the real valued sketch of size $2$ ($m=1$) computed over the dataset $\mathcal{X}$ as in (\ref{Eqn: The sketch}). It is possible to recover an estimate of the single depth location parameter $t_1$ directly from the sketch, without recourse to the data $\mathcal{X}$, via the trigonometric sample mean 
\begin{equation}
\label{eqn: motivation estim}
    \hat{t}_1 = \frac{T}{2 \pi} \phase \left\{ \sum_{j=1}^n \cos \left(\frac{2\pi x_j}{T}\right) + {\rm i}\sum_{j=1}^n \sin \left(\frac{2\pi x_j}{T} \right)\right\}
\end{equation}
\noindent where  $\phase$ denotes the phasor angle. As the background photons are distributed uniformly over the interval $[0,T-1]$ ($\pi_b(x)=\frac{1}{T}$), the expected moment of the photons originating from background sources is zero, $\E_{x\sim\pi_b}\Phi(x)=\mathbf{0}$. The resulting estimate of the single depth parameter $\hat{t}_1$ is therefore an unbiased estimator of the location parameter $t_1$. The estimator in (\ref{eqn: motivation estim}) coincides with the circular mean estimator detailed in \cite{jammalamadaka2001topics}.  Here the circular mean requires the first (non-zero) frequency. 
\par
\begin{figure}[ht!]
\centering
\includegraphics[scale=0.25]{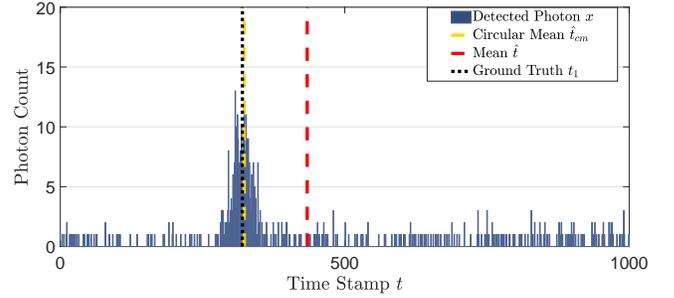}
\caption{The TCSPC histogram with $t_1=320$. The circular mean estimate (yellow) and the standard mean estimate (red) superimposed. }
\label{Fig: motivational hist}
\end{figure}

  Throughout the paper, we consider the detection point SBR, defined as $\frac{\sum_{k=1}^K\alpha_k}{\alpha_0}$, and not the raw sensor SBR which can be much lower in practice \cite{detectSBR}. We summarise the above using a simulated example, where a pixel of $T=1000$ histogram bins with a detection point signal-to-background ratio (SBR) of 1  and a total of $n=600$ photons is simulated, where the time-stamp of each photon is denoted by $\mathcal{X}=\{x_i\}^n_{i=1}$. The data was simulated using a Gaussian impulse response function with $\sigma=15$ and a true position at time-stamp $t_1=320$. Computing the sketch $\z_n$ from (\ref{Eqn: The sketch}) and using (\ref{eqn: motivation estim}) we obtain the sketch estimate  $\hat{t}_{\text{cm}}=323.3$ and the sample mean estimate of $\hat{t}=434.1$. The TCSPC histogram along with both the circular and standard mean estimates as well as the location parameter $t_1$ are shown in Figure \ref{Fig: motivational hist} where it is evident that the circular mean estimate does not suffer from the noise bias inherent in the sample mean.

\par Importantly, the sketch formed using the moment in (\ref{eqn: motivate Phi}) is equivalent to the complex valued ECF sketch $\z_n=\frac{1}{n}\sum^n_{j=1}e^{{\rm i}\omega x_j}$ sampled at $\omega=\frac{2\pi}{T}$, decoupled into both its real and imaginary components. In fact, the estimate $\hat{t}_1$ in (\ref{eqn: motivation estim}) is the optimal estimator to the compressive ECF sketch detailed in (\ref{Eqn: CL loss function}) (see Appendix \ref{Appendix: Circ mean}). Principally, we only need to store and transfer 2 values to accurately estimate the depth location of the object or surface, without the requirement to recourse to the original photon time-stamped data. For the remainder of this section, we generalize the approach of forming a sketch in (\ref{Eqn: The sketch}) of arbitrary size and sampling the ECF at multiple frequencies $\left [{\omega_i}\right ]^m_{i=1}$. This will enable us to obtain statistically efficient estimates for the single surface case and to solve more complex lidar scenes including several surfaces with varying intensities where more salient information of the observation model is required.

\subsection{Sampling the ECF}\label{Subsec: Sampling the ECF}
Recall that the observation model $\pi$ in (\ref{Eqn: Alternative mixture Observ model}) is discretized over the interval $[0,T-1]$ which we can consider to be a sufficient sampling if the distribution in (\ref{Eqn: Alternative mixture Observ model}) is approximately bandlimited. As a result, the characteristic function $\Psi_\pi(\omega)$ has a finite basis characterized by the set of frequencies
\begin{equation}
    \label{Eqn: Char fun finite basis}
    \mleft\{\frac{2\pi j}{T} \;\middle|\;  j=0,1,\dots,T-1 \mright\}.
\end{equation}
We can generalise the approach from Section \ref{subsec: warm up} by sampling multiple frequencies from the finite basis in order to construct the ECF sketch. As is the case for the circular mean, the frequencies $\omega=\frac{2\pi j}{T}$ for  $j=1,2,\dots,T-1 $  correspond to the zeros in the Dirichlet kernel function associated with the background pdf $\pi_b$ seen in (\ref{Eqn: Char function Obs Model}). We can therefore construct a sketch of arbitrary dimension $m$ that is also \textit{blind} to photons originating from background sources by avoiding the zero frequency $\omega=0$ of the finite basis. As a result, we define the set of orthogonal frequencies by
\begin{equation}
    \label{eqn: Orth Set of weights}
    \Omega \coloneqq \mleft\{\omega_j=\frac{2\pi j}{T} \;\middle|\;  j=1,2,\dots,T-1  \mright\}.
\end{equation}
We coin this set the \textit{orthogonal frequencies} as it defines regions over the interval of the observation model's characteristic function where the signal's contribution is orthogonal to the background's contribution.

\subsubsection{Sampling Schemes}\label{subsec sampling schemes}
In order to construct a sketch, we are ultimately interested in retaining sufficient salient information of the characteristic function $\Psi_\pi$ such that we can identify and estimate the unique location and intensity parameters $\theta$ of the observation model $\pi(x\mid\theta)$ defined in (\ref{Eqn: Alternative mixture Observ model}). It was discussed in Section \ref{Sec: Background} that the CF of a probability distribution decays in frequency, i.e. $\Psi_\pi(\omega)\rightarrow 0$ as $\omega \rightarrow \infty$. Furthermore, as the observation model is discretized over the interval, we assume that the characteristic function of the observation model is approximately band-limited. A natural sampling scheme would therefore be to sample the first $m$ frequencies of the orthogonal frequencies $\Omega$ to capture the maximum energy of the CF. In other words, we could truncate the CF of the observation model whilst avoiding the zero frequency. 
\par Alternatively, in \cite{gribonval2020statistical,keriven2018sketching}, provable guarantees for estimating mixture of Gaussian models have been provided,  under certain conditions based on random sampling (cf. compressive sensing \cite{eldar_kutyniok_2012}) of the CF. It is understood that the higher frequencies of the CF may provide further information to help discriminate distributions that are close in probability space. Moreover, if the CF decays slowly in frequency then the energy of the CF will be spread more throughout the set of orthogonal frequencies. We therefore provide an alternative sampling scheme whereby we randomly sample the set of orthogonal frequencies with respect to some sampling law $\Lambda$. In a similar design to the frequency sampling pattern proposed in \cite{keriven_conference}, we sample the orthogonal frequencies by
\begin{equation}
    (\omega_1,\omega_2,\dots,\omega_m)\sim \Lambda_{\hat{h}},
\end{equation}
where $\Lambda_{\hat{h}}\propto\hat{h}$. To formalize, we consider the follow sampling schemes in order to construct our ECF sketches:

\begin{enumerate}
    \item Truncated Orthogonal Sampling: Sample the first $m$ frequencies i.e $j=1,2,\dots,m$ from $\Omega$.
    \item Random Orthogonal Sampling:  Sample the set of frequencies randomly, governed by the distributing law $\Lambda_{\hat{h}}$.
\end{enumerate}
Depending on the circumstances of the lidar device we might expect one or the other sampling scheme to perform better.

 \subsection{Practical Hardware Considerations}
\subsubsection{Online Processing}
One of the major advantages of forming a sketch $\mathbf{z}_n$, as in (\ref{Eqn: The sketch}), is that it is naturally amenable to online processing. Recall that for an arbitrary pixel in the scene, the resulting sketch that can be transferred off-chip is $\z_n=\sum^n_{i=1}\Phi(x_i)$. Algorithm \ref{Alg: Online proccessing} demonstrates how the sketch for a given pixel is updated in real time during an acquisition window where $n$ photons are detected by the SPAD array. For each photon arrival $x_j$ during the acquisition window, an intermediate sketch is accumulated as well as an integer counter. Once the acquisition window is over, the resulting sketch is transferred off-chip for post-processing.

\begin{algorithm}
\caption{Sketch Online Processing}
\label{Alg: Online proccessing}
\begin{algorithmic}
\STATE \textbf{Initialisation:} $\mathbf{z}=0, n=0$
\WHILE{Acquisition Window}
\IF{New Photon Arrival $x_j$}
\STATE $\mathbf{z}\xleftarrow{} \mathbf{z} + \Phi(x_j)$
\STATE $n\xleftarrow{}n+1$
\ENDIF
\ENDWHILE
\STATE $\mathbf{z} \xleftarrow{}\mathbf{z}/n$
\ENSURE The sketch $\z$ is transferred off-chip for post-processing.
\end{algorithmic}
\end{algorithm}

\par This is very beneficial as all that is needed to be stored on-chip is the sketch $\z$ of size $2m$ and an integer counter. As such, forming the sketch in an online processing manner, as in Algorithm \ref{Alg: Online proccessing}, circumvents the need to compute and store a large histogram or store each individual photon time-stamp. Moreover, it should be noted that no further hardware is required to form the sketch and existing lidar devices can be easily adapted to implement our proposed technique.
\par The computation of the sketch itself requires the calculation of the Fourier features i.e. $\cos(2\pi\omega_j/T)$ and $\sin(2\pi\omega_j/T)$, which would have to be computed in real time for each time-stamp. However, various efficient logic-based schemes already exist for performing such computations \cite{cordic_ref} based on either the classic CORDIC algorithms or polynomial approximations. Alternatively, in \cite{schellekens2021asymmetric}, Schellekens et al. show that in principle one can also replace the Fourier features by alternative periodic functions (e.g. square waves or triangle waves) in conjunction with random dithering. Subsequently, we will assume that we have access to sufficiently accurate sketch values and leave details of specific hardware implementation for future work.

\section{Sketched Lidar Reconstruction}  \label{Sec: Sketched Lidar Reconstruction}
\subsection{Statistical Estimation}
Once the ECF sketch is constructed using either sampling scheme, we must estimate the parameters $\theta$ of the observation model $\pi(x\mid\theta)$ solely from the sketch $\z_n$. In general, there is no closed form expression to estimate $\theta$ from the sketch of arbitrary size as is the case for the circular mean estimate in (\ref{eqn: motivation estim}). It is well documented in the ECF and GeMM literature, e.g. \cite{10.2307/2958763,10.2307/1912775,gemmhall}, that a complex valued ECF sketch $\z_n$ of size $m$, computed over a finite dataset $\mathcal{X}=\{x_1,\dots,x_n\}$, satisfies the central limit theorem. Formally, a sketch $\z_n\in\mathbb{C}^m$ converges asymptotically to a Gaussian random variable
\begin{equation}
\label{Eqn: sketch central limit theorem}
    \z_n\xrightarrow[]{\text{dist}}\mathcal{N}\big(\Psi_\pi,n^{-1}\Sigma_\theta\big),
\end{equation}
where $\Sigma_\theta\in\mathbb{C}^{m\times m}$ has entries $(\Sigma_\theta)_{ij}=\Psi_\pi(\omega_i-\omega_j)-\Psi_\pi(\omega_i)\Psi_\pi(-\omega_j)$ for $i,j=1,2,\dots,m$. The asymptotic normality result in (\ref{Eqn: sketch central limit theorem}) naturally leads to a sketch maximum likelihood estimation (SMLE) algorithm that consists of minimising the following 
\begin{equation}
    \label{Eqn: SMLE Scheme}
    \argmin_\theta\,\, \frac{m}{2}\log\det(\Sigma_\theta)+n(\z_n-\z_\theta)^T\Sigma_\theta^{-1}(\z_n-\z_\theta),
\end{equation}
where for convenience we denote
$\z_\theta=\left[\Psi_\pi(\omega_j)\right]_{j=1}^m$. For an observation model consisting of $K$ surfaces and a general impulse response function $h$, recall that 
\begin{equation}
    \label{eqn: general sketch analytic}
    \z_\theta=\left[\sum^K_{k=1}\alpha_k\hat{h}(\omega_j)e^{{\rm i}\omega_jt_k}\right]_{j=1}^m
\end{equation} and $\theta=(\alpha_0,\alpha_1,\dots,\alpha_K,t_1,\dots,t_K)$. Note that we have dropped the Dirichlet kernel function on the assumption that we are using one of the proposed sampling schemes. Minimising (\ref{Eqn: SMLE Scheme}) is equivalent to minimising the compressive GeMM objective function defined in (\ref{Eqn: CL loss function}) with the weighting function chosen to be $\mathbf{W}=\Sigma_\theta^{-1}$. The weighting matrix $\mathbf{W}=\Sigma_\theta^{-1}$ is asymptotically optimal in the sense that it minimises the variance of the estimator $\hat{\theta}$ from the sketch $\z_n$ \cite{gemmhall}. 
\par In practice $\Sigma_\theta$ is $\theta$ dependent as it is a function of the underlying parameters $\theta$ that are to be estimated. There are various well established methods in the GeMM and ECF literature \cite{hansenGeMM,gemmhall} that tackle the difficulty of approximating $\Sigma_\theta$ and estimating $\theta$ simultaneously. In \cite{osomeFouriermethods}, they use the K-L method which iteratively estimates $\Sigma_\theta$ and $\theta$ in a two stage procedure by fixing and updating one at a time, resulting in a computation complexity of $\mathcal{O}\left(m^3\right)$ due to inverting $\Sigma_\theta$. Some particular methods \cite{hansen2step} fix $\Sigma_\theta$ after only a few iterations of the K-L approach to reduce the computational complexity of the algorithm, although this typically comes at the cost of introducing sample bias \cite{HAUSMAN201145}. Occasionally, the covariance matrix is set throughout to be the identity, $\Sigma_\theta=I$, reducing (\ref{Eqn: SMLE Scheme}) to a standard least squares optimization and a computational complexity of $\mathcal{O}\left(m\right)$, however this generally results in a less statistically efficient estimator $\hat{\theta}$\cite{hansenGeMM}. In this paper, we estimate $\Sigma_\theta$ and $\theta$ simultaneously at each iteration. This approach is commonly referred to as Continuous Updating Estimator  (CUE) \cite{hansen2step} and obtains estimates that do not produce sample bias like the two-step K-L approach \cite{HAUSMAN201145} and can often lead to more statistically efficient estimators \cite{hansenGeMM}. However, the SMLE method is not restricted to the CUE and in certain situations practitioners may choose to sacrifice unbiased and efficiently optimal estimators for a reduced computational complexity by considering the other methods discussed.
\par The optimisation problem in (\ref{Eqn: SMLE Scheme}) is also typically non convex and can suffer from spurious local minima. For the case when there is only a single surface, we initialise the SMLE algorithm using the analytic circular mean solution in (\ref{eqn: motivation estim})  with minimal added computational overhead. From our experience with synthetic and real data, the circular mean estimate generally initialises the SMLE algorithm within the basin of the global minima, hence the issues associated with non-convex optimization are circumvented. For the case of multiple surfaces, we form a coarse uniform grid across $[0,T-1]^K$ and initialise at the smallest SMLE loss.

\par In the orthogonal sampling scheme, one could alternatively zero-pad the sketches, perform an inverse FFT (iFFT) and find the maximum peak to estimate the depth position of the surface. However, this approach is fundamentally different to that of the orthogonal truncated sketch as the iFFT method is simply a low pass approximation of the TCSPC histogram whereas the proposed SMLE algorithm performs nonlinear parameter fitting. As a result, the iFFT method will be particularly inaccurate at distinguishing between closely spaced reflectors. In contrast to the proposed sketched lidar acquisition, the iFFT method does not take into account the particular nature of the IRF and achieves poor depth accuracy in the presence of a non-symmetric IRF (see Appendix \ref{appendix: comparison to transient imag}). Furthermore, the iFFT approach requires $\mathcal{O}(T)$ off-chip memory complexity in comparison to $\mathcal{O}(m)$ of our proposed SMLE algorithm.
\subsection{Central Limit Theorem} \label{subsec: CLT}
\begin{figure}[ht!]
\centering
\includegraphics[width=1\linewidth]{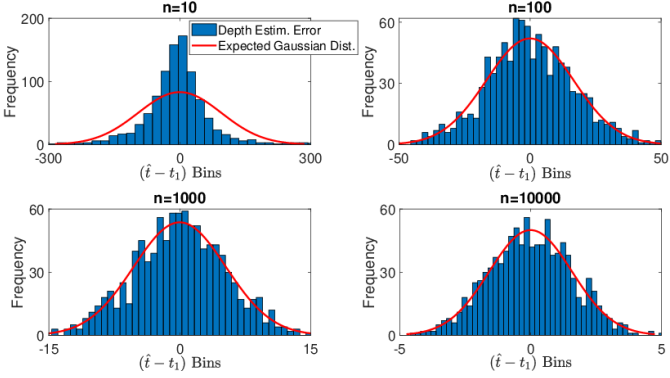}
\caption{ Histograms of the estimation error $(\hat{t}-t_1)$ for increasing photon count $n$ where the sketched lidar estimate (circular mean) is denoted by $\hat{t}$. The expected error distribution in (\ref{Eqn: sketch central limit theorem}) is depicted in red.}
\label{fig: CLT hists}
\end{figure}

One of the main advantages of the SMLE lidar approach from (\ref{Eqn: sketch central limit theorem}) is that even at low photon levels (i.e. small $n$), the SMLE estimates quickly follow the central limit theorem (CLT) and provide a good approximation of its expectation. In contrast, the TCSPC histogram used for many estimation methods, discussed in Section \ref{Sec: Intro}, is a poor approximation to its expectation as each time-stamp bin $t$ has only a small number of photons. Thus efficient processing of the full histogram data requires careful consideration of the underlying Poisson statistics \cite{VivekLidar}. This is illustrated in Figure \ref{fig: CLT hists} which shows four separate histograms of the error $(\hat{t}-t_1)$ for increasing photon count $n$, along with the asymptotic Gaussian distribution from (\ref{Eqn: sketch central limit theorem}). The estimate $\hat{t}$ was obtained from a real valued sketch of size 2 ($m=1$) using the circular mean estimate in (\ref{eqn: motivation estim}). The simulated data was the same as the motivation example in Section \ref{subsec: warm up} where a Gaussian IRF with $\sigma=15$ was used. The SBR was set at 1 and the total number of time-stamps was $T=1000$. The total photon count varied from $n=10$ to $n=10000$ increasing by a factor of 10 each time. For each photon count, we estimated the location parameter $t_1$ a total of 1000 times where the data $\mathcal{X}=\{x_i\}_{i=1}^m$ was simulated independently for each trial.
\par Even at extremely low photon counts of $n=10$, the error $(\hat{t}-t_1)$ can be reasonably approximated by a Gaussian random variable centred around 0. This suggests that the estimate $\hat{t}$ quickly satisfies the central limit theorem with respect to the photon count $n$. Further analysis of the proposed SMLE algorithm in the photon starved regime can be seen in Appendix \ref{App: Photon Starved Regime}. In the large photon regime ($n=10000$), the estimation error is concentrated tightly around zero and mostly contained within 5 time-stamps. These results suggest that the sketched lidar CLT results of (\ref{Eqn: sketch central limit theorem}) hold even for low photons levels, hence the SMLE loss in (\ref{Eqn: SMLE Scheme}) is a well-justified loss to minimise.  A further potential benefit from this asymptotic normality is that it permits us to directly use \textit{plug-and-play} Gaussian denoising algorithms to further improve the imaging performance \cite{tachellaNComms,RappPhotonEfficient}.   

\subsection{Statistical Efficiency}\label{Subsec: Statistical Efficiency}
In this section, we calculate the theoretical statistical efficiency of the sketched lidar estimates, $\theta$, that parametrize the observation model $\pi(x\mid\theta)$ in (\ref{Eqn: Alternative mixture Observ model}), and compare them with the estimates obtained using the full data (i.e no compression) using the relative error percentage. The relative error percentage, which will be defined later, is a key metric allowing us to quantify the relative loss of information given a sketch of size $m$ from a statistical point of view.
\par Statistical efficiency is a measure of the variability or quality of an unbiased estimator $\hat{\theta}$ \cite{FisherEfficiency}. The Cram\'er-Rao bound gives a lower bound on the mean squared error of $\hat{\theta}$ \cite{amari2007methods} and therefore provides a best case scenario on the variability of the parameter estimates. Given the observation model $\pi(x\mid\theta)$ and the corresponding Fisher information matrix (FIM), defined as 
\begin{equation}
\label{eqn: Fisher info matrix}
    \mathcal{I}_{\text{data}}(\theta)\coloneqq n \E \Bigg[\Bigg(\frac{\partial \log \pi(x \mid \theta)}{\partial\theta}\Bigg)^2\Bigg],
\end{equation}
then the optimal Cram\'er-Rao mean squared error, in terms of the full data, is defined as 
\begin{equation}
    \label{eqn: Optimal MSE}
    \text{RMSE}_n\coloneqq\sqrt{\sum^{2K}_{k=1} [\mathcal{I}_{\text{data}}(\theta)^{-1}]_{\{kk\}}}.
\end{equation}
Equivalently, we can compute the FIM for the sketched case using the normality result stated in (\ref{Eqn: sketch central limit theorem}), where the FIM of a multivariate Gaussian distribution \cite{amari2007methods} is defined as 
\begin{equation}
\label{Eqn: Sketched FIM}
\big(\mathcal{I}_{\text{sketch}}(\theta)\big)_{ij}\coloneqq n\dfrac{\partial\z_\theta}{\partial\theta_i}\Sigma^{-1}_{\theta_0}\dfrac{\partial\z_\theta}{\partial\theta_j},
\end{equation}
where $\z_\theta$ is the sketch defined in (\ref{eqn: general sketch analytic}). Similarly, we define the optimal sketched Cram\'er-Rao mean squared error as 
\begin{equation}
    \label{eqn: Optimal Sketch MSE}
    \text{RMSE}_m\coloneqq\sqrt{\sum^{2K}_{k=1} [\mathcal{I}_{\text{sketch}}(\theta)^{-1}]_{\{kk\}}}.
\end{equation}
\noindent To quantify the statistical efficiency of an estimate obtained from a real valued sketch of size $2m$, we use the relative error percentage (REP) metric which compares the optimal sketch root mean squared error $\text{RMSE}_m$ with the corresponding full data root mean squared error $\text{RMSE}_n$, defined by
\begin{equation}
   \text{REP}\coloneqq100\Bigg(\frac{\text{RMSE}_m-\text{RMSE}_n}{\text{RMSE}_n}\Bigg).
\end{equation}
Notably, the FIM of the sketched statistic in (\ref{Eqn: Sketched FIM}) scales with $n$, hence the REP metric is independent of the photon count. We compare the statistical efficiency of the sketched lidar estimates to the alternative compression technique of coarse binning \cite{hardware} discussed in Section \ref{Sec: Intro}. The coarse binning approach can be seen to be equivalent to constructing a summary statistic
\begin{equation}
    \label{eqn: Coarse bin sketch}
   \Tilde{\z}_n =\sum^n_{i=1}\left\{\mathbbm{1}_{[(j-1)\Delta_{\tilde{m}},j\Delta_{\tilde{m}}]}(x_i)\right\}_{j=1}^{\tilde{m}},
\end{equation}
where $\Delta_{\tilde{m}}=\big\lceil\frac{T}{\tilde{m}}\big\rceil$ denotes the down-sampling factor, $\tilde{m}$ denotes the number of measurements equivalent to the real-valued sketch size (i.e. $\tilde{m}=2m$)  and $\mathbbm{1}_{[t_i,t_i+\Delta_{\tilde{m}}]}(x)$ is the indicator function defined as
\begin{equation}
    \label{eqn: Indicator function Coarse}
    \mathbbm{1}_{[(j-1)\Delta_{\tilde{m}},j\Delta_{\tilde{m}}]}(x)\coloneqq 
    \begin{cases}
   1 ~&\text{ if }~ x\in {\big[(j-1)\Delta_{\tilde{m}},j\Delta_{\tilde{m}}\big)},\\
   0 ~&\text{ Otherwise. }
    \end{cases}
\end{equation}
Once the coarse binning sketch has been constructed, traditional estimation methods, for e.g matched filtering \cite{logmatchedfilter} or expectation maximization \cite{EMAlg}, can be employed to estimate the parameters of the observation model.
\par
\begin{figure}
\centering
\includegraphics[width=0.48\textwidth]{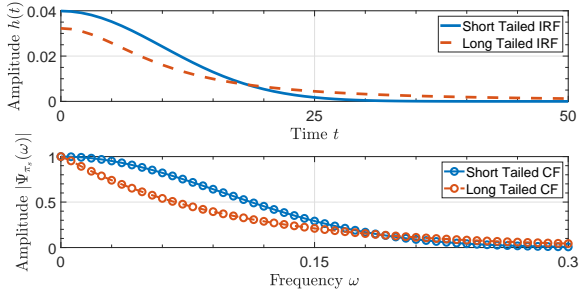}
\caption{ The CF (top) of a short (blue solid) and long (red dashed) tailed impulse response function (bottom). }
\label{Fig: eff two tailed}
\end{figure}

  Lidar scenes typically have only 0, 1 or 2 reflectors in the scene, although in some specific applications, for example airborne lidar \cite{LidarFoliage}, tree-canopy foliage can return $K>2$ reflectors. Our proposed method can handle greater number of reflections, however in the following experiments we only consider the typical case where $K=1,2$.      Moreover, we choose the setting of the lidar scene (e.g. binning resolution, peak location, intensity) to best replicate a realistic setting as seen in Section \ref{Subsec: Real Data}. In each experiment, we consider two different impulse response functions (IRF), exhibiting both a short and long-tail. Figure \ref{Fig: eff two tailed} depicts the contrasting IRFs and the magnitude of their corresponding characteristic functions, $\Psi_{\pi_s}(\omega)=\hat{h}(\omega)e^{{\rm i}\omega t}$. We evaluate the statistical efficiency of the sketched and coarse binning estimate using the REP as a function of the number of real measurements $2m$ and examine both the random and truncated orthogonal sampling schemes discussed in Section \ref{subsec sampling schemes}. 

\subsubsection{One Surface}
\begin{figure}
\centering
\includegraphics[width=0.48\textwidth]{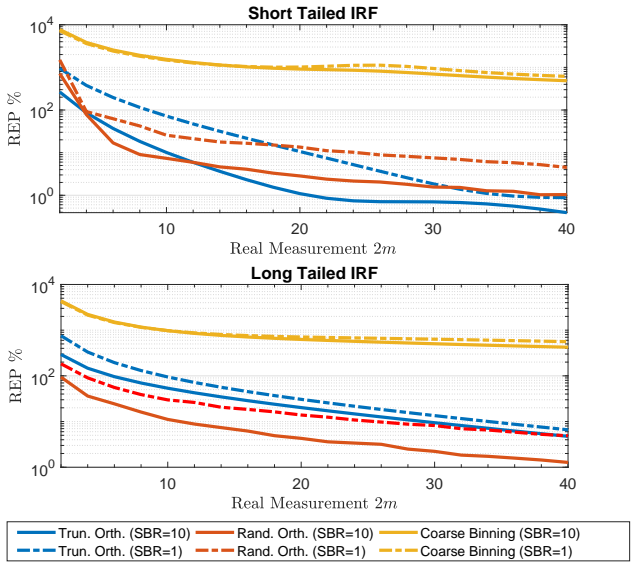}
\caption{The REP as a function of the number of real measurements ($2m$) for a single peak lidar scene.}
\label{Fig: Unimodal Efficiency}
\end{figure}

We first evaluate the REP for a single peak case positioned at $t_1=430$, a window size of $T=1000$. We consider both low and high background photon count levels, where the SBR was set at 10 and 1, respectively. Figure \ref{Fig: Unimodal Efficiency} shows the REP metric as a function of the number of real measurements $2m$ for the truncated orthogonal (blue), random orthogonal (red) and coarse binning (orange) compression techniques, where the high (SBR=10) and low (SBR=1) background photon levels are denoted by a solid and dashed line, respectively. The top and bottom plots depict the short and long-tailed IRF, accordingly. We first observe that both sketched lidar sampling schemes approach $0\%$ as the real measurements increase and only a modest number of measurements is needed to obtain a low REP. In contrast, the coarse binning approach exhibits a slow convergence REP and remains high throughout the measurement range. Importantly, we see that the different sketch sampling schemes outperform each other depending on the tail of the IRF and hence the rate of decay of the CF. For instance, the truncated scheme produces a lower REP for the short-tailed IRF, while the random sampling scheme achieves a quicker convergence and a significantly lower REP throughout the measurement range for the long-tailed IRF. This can be explained by Figure \ref{Fig: eff two tailed}, the CF of the short-tailed IRF has the majority of its energy contained within the first few ($m=10$) frequencies, while the CF of the long-tailed IRF has its energy spread more throughout its frequency.

\subsubsection{Two Surfaces}
\begin{figure}
\centering
\includegraphics[width=0.48\textwidth]{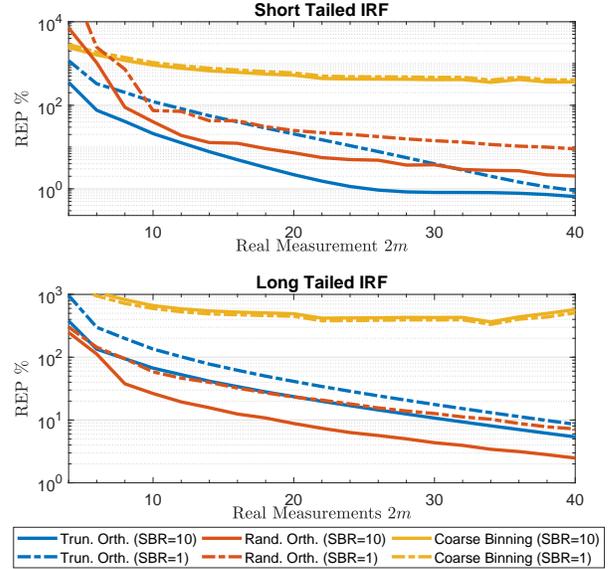}
\caption{The REP as a function of the number of real measurements ($2m$) for a lidar scene with 2 surfaces.}
\label{Fig: Bimodal Efficiency}
\end{figure}

We now evaluate the REP for a two peak case positioned at $(t_1,t_2)=(320,570)$, a window size of $T=1000$. The intensity of the two peaks is given by $75\%$ and $25\%$, respectively, simulating an object that is positioned behind a semi-transparent surface. We simulate both low and high background levels, where the SBR was again set at 10 and 1, respectively. Figure \ref{Fig: Bimodal Efficiency} shows the REP metric as a function of the number of real measurements $2m$ for the truncated orthogonal (blue), random orthogonal (red) and coarse binning (orange) compression techniques, where the high (SBR=10) and low (SBR=1) background photon levels are denoted by a solid and dashed line, respectively. The top and bottom plots depict the short and long-tailed IRF, accordingly. We see the same pattern as the single surface case where the REP remains high for the coarse binning compression technique while, in contrast, the sketched lidar converges towards a relatively low REP in a modest number of measurements. We again observe that the truncated scheme performs best on a fast decaying CF, while the random sampling scheme outperforms the truncated counterpart when there is a slow decaying CF.  The doubling of the dimension of the parameter $\theta$ by estimating two peaks and intensities, does not have a significant impact on the required number of measurements needed to achieve a relatively low REP. For instance in the high SBR (solid) scenario, the truncated orthogonal sampling scheme requires $20$ real measurements ($m=10$) to achieve a REP less than $1\%$  for the unimodal case compared with a requirement of $24$ real measurements ($m=12$) to achieve the same level of REP for the bimodal case. These theoretical results on the statistical efficiency of the lidar sketch show that only a moderate sketch size is needed to achieve negligible loss of information. The results are based on the asymptotic normality property discussed in (\ref{Eqn: sketch central limit theorem}), and we have seen in Section \ref{Subsec: Syn Data} that in practice this normality result holds even for small photon counts of $n=10$.

\par In coarse binning, it can be beneficial to broaden the impulse response (while keeping laser power constant) such that it covers more than a single coarse bin. This strategy can achieve (coarse) sub-bin resolution (see for example \cite{Gyongy:20}). Furthermore, Gyongy et al. \cite{Gyongy:20} proposed an algorithm that estimates the depth position continuously, in contrast to quantization limited matched filtering \cite{logmatchedfilter}. We further compare our proposed sketched lidar method to the wide pulse width coarse binning and algorithm used in \cite{Gyongy:20} for a range of SBR values. In the simulation, the photon count was set at $n=100$ and a Gaussian IRF was used. For the wide pulse width, we replicate the lidar device by setting $\sigma_1=0.4$. To compare with the narrow pulse width settings, we set  $\sigma_2=5$. In both scenarios, a total of $2m=16$ coarse bins are used. For our proposed SMLE algorithm, we compare the same compression be taking a (real-valued) sketch of size 16 ($m=8$). We evaluate the depth estimation over SBR values ranging between $10^{-1}$ to $10^2$  for 250 Monte-Carlo simulations. The coarse binning CRB is calculated where the best pulse width has been optimally selected for each SBR level.

\begin{figure}[ht!]
\centering
\includegraphics[width=1\linewidth]{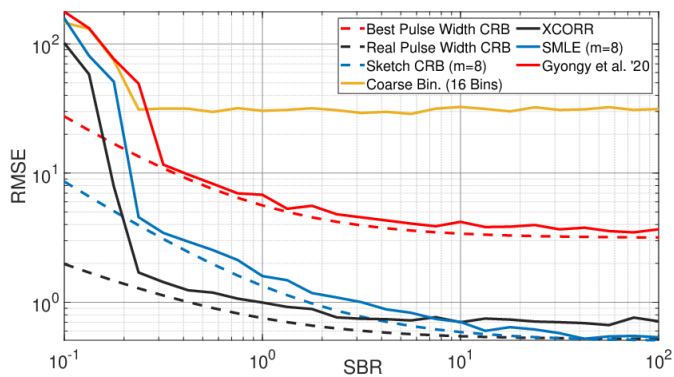}
\caption{Comparison of the RMSE achieved by  wide and narrow Gaussian pulse width coarse binning to our proposed SMLE algorithm.}
\label{Fig: narrow wide width compare}
\end{figure}
 
\noindent As shown in Figure \ref{Fig: narrow wide width compare}, the coarse sub-bin resolution can indeed improve the resolution with respect to coarse binning in large SBR regimes, but it still falls significantly behind the resolution obtained using the narrowest IRF with a fine scale time-stamp of our proposed sketch method. For instance, at an SBR of 0.23 the wide pulse width achieves a RMSE of 264.6 bins compared to 31.1 and 4.5 bins for the narrow pulse width coarse binning and SMLE, respectively. As the pulse width optimised algorithm in \cite{Gyongy:20} only exhibits significant improvement in the high SBR scenario we do not consider it further in the paper.

\section{Experiments}\label{Sec: Experiments}
\subsection{Experimental set up}
In this section, we evaluate our compressive lidar framework on synthetic and real data with increasingly complex scenes. Our method is compared with classical algorithms working on the full data space (i.e no compression) namely matched filtering \cite{logmatchedfilter} and expectation maximization (EM) \cite{EMAlg}. Moreover, we also compare our results to the alternative compression technique of coarse binning \cite{hardware} discussed in Section \ref{Sec: Intro} and (\ref{eqn: Coarse bin sketch}). Both the matched filtering and EM algorithms estimate the location parameters using the full data and therefore the results obtained from these methods set a benchmark to the estimation accuracy when no compression takes place.  For sake of fair comparison, we use the real valued sketch in all the subsequent results, such that the number of real measurements is equivalent to $2m$.
\subsubsection{Processing}
Restoration of depth imaging of single-photon lidar consists of estimating a 3D point cloud from a lidar data cube containing the number of photons $n_{i,j,t}$ in pixel $(i,j)$ at time-stamp $t$, where $i=1,2,\dots,N_r, j=1,2\dots,N_c$  and $t = 0,1,\dots,T-1$. We denote the average photon count for each pixel by $\bar{n}$ and process each pixel $(i,j)$ of the data cube and estimate the true location and intensity parameter, denoted $t_1$ and $\alpha$, respectively.  The intensity of a point in pixel $(i,j)$ of the point cloud is calculated by the number of photons in the pixel multiplied by the proportion of the signal i.e. $\alpha_k \sum_{t=0}^{T-1}n_{i,j,t}$.  A data driven impulse response is given for each dataset and we can obtain the characteristic function of the IRF by using (\ref{Eqn: Char function Obs Model}). 
\subsubsection{Evaluation Metrics}
Two different error metrics are used to evaluate the performance of our proposed sketched lidar framework. We consider the root mean squared error (RMSE) between the reconstructed image and the ground truth. Given that $t_{i,j,k}$ is the location of the $k$th peak in pixel $(i,j)$ and $\hat{t}_{i,j,k}$ the estimated counterpart, then the root mean squared error of the reconstructed image is
\begin{equation}
    \text{RMSE} := \sqrt{\frac{1}{KN_rN_c}\sum_{i=1}^{N_r}\sum_{j=1}^{N_c}\sum^K_{k=1}\Big( t_{i,j,k}-\hat{t}_{i,j,k}\Big)^2}.
\end{equation}
The compression of both the sketched lidar and coarse binning approach is measured in terms of the dimension reduction achieved by the statistic with respect to the raw TCSPC data and is quantified by the metric $\max \{\frac{2m}{T},\frac{2m}{n}\}$, which is dependent on the dimensions, $T$ and $n$, of the lidar scene and where the number of real measurements ($2m$) is used for sake of fair comparison.   

\subsection{Synthetic Data}\label{Subsec: Syn Data}

\begin{figure}[ht!]
\centering
\includegraphics[width=1\linewidth]{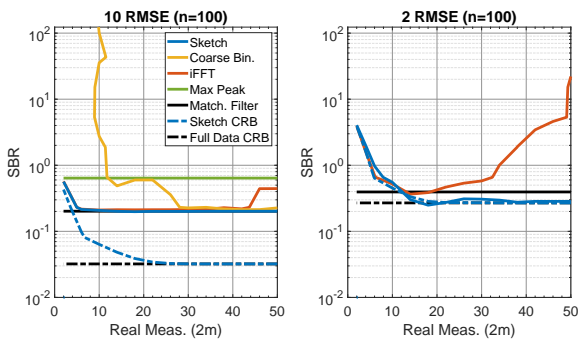}
\caption{RMSE level set contour plots for varying SBR levels and number of real measurements $2m$ for a photon count of $n=100$. The RMSE level are $10\Delta \tau$ (left) and $2\Delta \tau$ (right). The legend is defined for both plots.}
\label{Fig: synthetic contour1}
\end{figure}

\begin{figure}[ht!]
\centering
\includegraphics[width=1\linewidth]{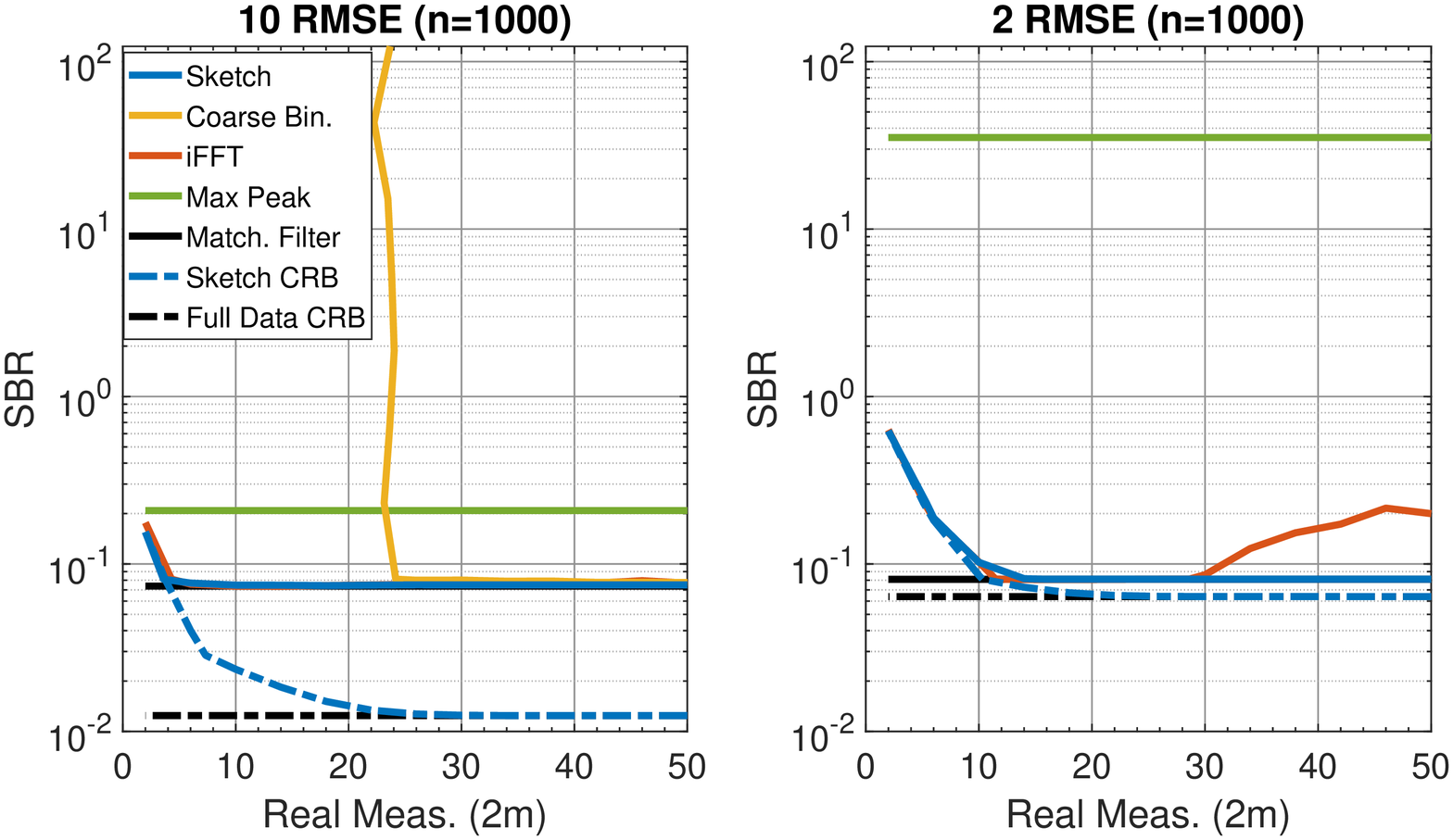}
\caption{RMSE level set contour plots for varying SBR levels and number of real measurements $2m$ for a photon count of $n=1000$. The RMSE level are $10\Delta \tau$ (left) and $2\Delta \tau$ (right). The legend is defined for both plots.}
\label{Fig: synthetic contour2}
\end{figure}

We evaluate the sketched lidar framework on a synthetic dataset simulating a pixel in a scene which consists of a single peak response. We chose the parameters that best replicated a realistic lidar scene and that were akin to the real datasets which will be discussed in \ref{Subsec: Real Data}. Therefore, we set the binning resolution at $T=250$, and impulse response was generated with a true Gaussian function where $\sigma=5$. We ran a Monte-Carlo simulation with $1000$ trials to evaluate and compare the performance of our sketched lidar framework for photon counts $n\in(100,1000)$ with varying SBR levels and number of real measurements $2m$. For each trial, we uniformly chose $t_1\sim\mathcal{U}(0,249)$, and estimated $\hat{t}$ for the sketched lidar approach, the iFFT method discussed in Section \ref{Sec: Sketched Lidar Reconstruction} as well the alternative compression technique of coarse binning. As a reference, we computed the matched filter estimate as well as estimating the maximum peak of the full histogram which represent the estimates over the full data (i.e. no compression). We varied the total number of real measurements between $2$ ($m=1$) and $50$ ($m=25$) and increased the SBR ratio from $10^{-2}$ to $10^2$ on a log-scale. Here we only show the results for the truncated orthogonal sampling scheme but we observed in practice that the alternative random orthogonal sampling scheme produces similar results. Figures \ref{Fig: synthetic contour1} and \ref{Fig: synthetic contour2} show the contour plots of the RMSE level of $10\Delta \tau$ (left) and $2\Delta \tau$ (right) for both $n=100$ and $n=1000$, respectively. The sketched lidar (solid blue), coarse binning (orange) and the iFFT (red) methods are depicted alongside the full data approaches of matched filtering (solid black) and maximum peak estimation (green). As discussed in Section \ref{Subsec: Statistical Efficiency}, the full data (dashed black) and the sketched (dashed blue) Cram\'er-Rao bound are given as reference and define the lower bound to the contour plot. Both the sketched lidar and iFFT approach converge quickly towards the full data estimate of matched filtering within 10 real measurements for both RMSE level sets and photon counts. In contrast, the coarse binning approach needs approximately 30 real measurements to achieve a similar performance as our sketched lidar method in achieving a RMSE of 10 bins. Moreover, coarse binning does not attain an RMSE of 2 for $2m\leq 50$ hence does not appear in the right subplot of Figure \ref{Fig: synthetic contour1}. It can be seen for a larger number of real measurements, the iFFT approach begins to diverge. This is because for larger number of measurements, the iFFT produces a less smooth linear approximation of the histogram and therefore it is more challenging to estimate the depth position.

\par
\begin{figure}[ht!]
\centering
\includegraphics[width=1\linewidth]{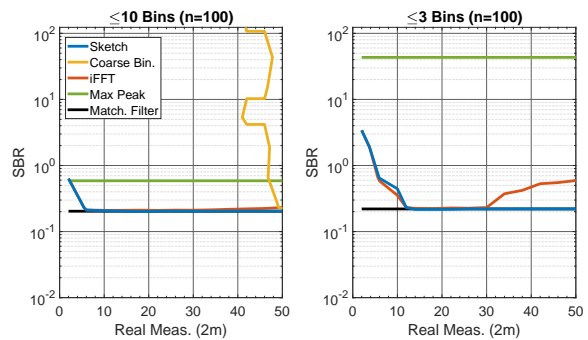}
\caption{RMSE level set contour plots for varying SBR levels and number of real measurements $2m$ for a photon count of $n=100$ for detecting $95\%$ of peaks within the level sets of $10\Delta \tau$ (left) and $3\Delta \tau$ (right). The legend is defined for both plots.}
\label{Fig: synthetic detection contour1}
\end{figure}

\begin{figure}[ht!]
\centering
\includegraphics[width=1\linewidth]{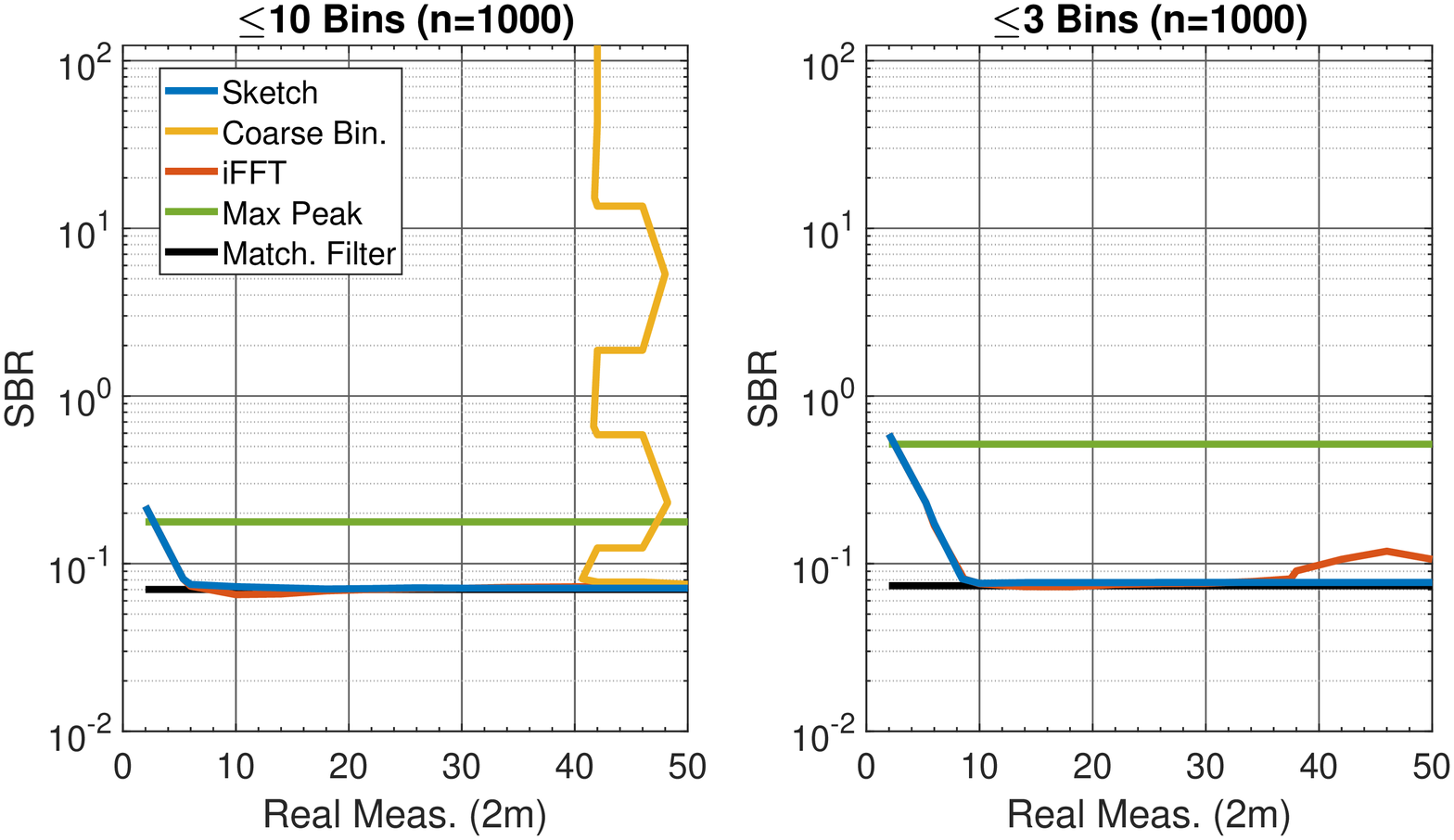}
\caption{RMSE level set contour plots for varying SBR levels and number of real measurements $2m$ for a photon count of $n=100$ for detecting $95\%$ of peaks within the level sets of $10\Delta \tau$ (left) and $3\Delta \tau$ (right). The legend is defined for both plots.}
\label{Fig: synthetic detection contour2}
\end{figure}

 Figures \ref{Fig: synthetic detection contour1} and \ref{Fig: synthetic detection contour2} show the $95\%$ of peaks detected within the level sets of $10\Delta \tau$  (left) and $3\Delta \tau$ (right). Our proposed sketch method achieves the same estimation performance as the full data matched filtering approach within approximately 12 real measurements ($m=6$) for all varying SBR ratios and photon counts. In contrast, the coarse binning approach requires approximately 45 real measurements, equating to a modest compression of $0.25$, to achieve $95\%$ of detections within $10\Delta \tau$. Furthermore, the coarse binning method could not achieve $95\%$ of detections within $3\Delta \tau$ for all the real measurements considered. These initial results on synthetic lidar data for a range of different SBR ratios and photon counts highlight the clear trade-off between compression and loss of temporal resolution for the coarse binning approach. In contrast, our proposed sketched lidar method overcomes the trade-off between compression and loss of resolution and only requires a very modest sketch size to achieve the same estimation performance as matched filtering using the whole data.

\subsection{Real Data}\label{Subsec: Real Data}
In this section we evaluate our sketched lidar framework on two real datasets of increasing complexity. Namely, a polystyrene head imaged at Heriot-Watt University \cite{altmann1,manipop} which consists mostly of a single peak, and a scene where two humans are standing behind a camouflage net, depicted in \cite{camo,tachellaNComms}, which contains of 2 objects per pixel with varying intensity.

\subsubsection{Polystyrene Head}\label{subsubsec: head dataset} The first scene consists of a polystyrene head placed 40 meters away from the lidar device. The data cube has width and height of 141 pixels, $N_r=N_c=141$ and a total of $T=4613$ time-stamps. A total acquisition time of 100 milliseconds was used for each pixel resulting in an average photon count of $\bar{n}=337$ with an SBR of approximately 6.82. The vast majority of pixels consist of a single peak, although there are a minority of pixels around the borders of the head that consist of two peaks. The parameter set to be estimated for each pixel is $\theta=(t,\alpha)$ of dimension 2. We compare our results with the ground truth obtained from the experiment as well as the full data algorithm of matched filtering and the coarse binning compression technique. As the matched filter is the maximum likelihood estimation of a single peak, we assume each pixel has one surface for the sake of comparison. As a result, we set the SMLE algorithm to estimate a single peak, however in practice we can use detection algorithms, for instance the sketch-based detection scheme proposed in \cite{sheehan2021surface}, to detect the number of surfaces present before estimation. The coarse binning approach is computed using  matched filtering once the data cube is down-sampled.

\begin{figure}[ht!]
\centering
\includegraphics[scale=0.25]{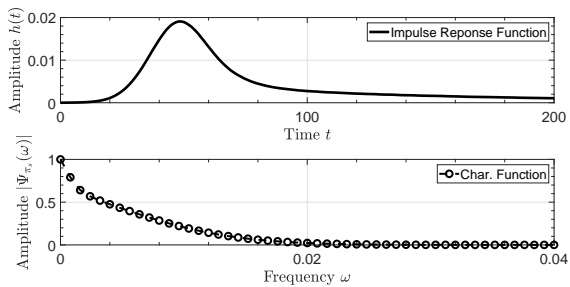}
\caption{ The CF (bottom) of the data driven impulse response function (top) of the polystyrene head dataset. }
\label{Fig: face char func}
\end{figure}

The data driven impulse response function and its corresponding CF obtained from (\ref{Eqn: Char function Obs Model}), are shown in Figure \ref{Fig: face char func}. We only present the results for the truncated orthogonal sampling scheme, from Section \ref{subsec sampling schemes}, but we observed in practice that the alternative random orthogonal sampling scheme produces similar results. We initialise the sketched lidar algorithm using the analytic circular mean solution in (\ref{eqn: motivation estim}).

\par
\begin{figure}[ht!]
\centering
\includegraphics[scale=0.25]{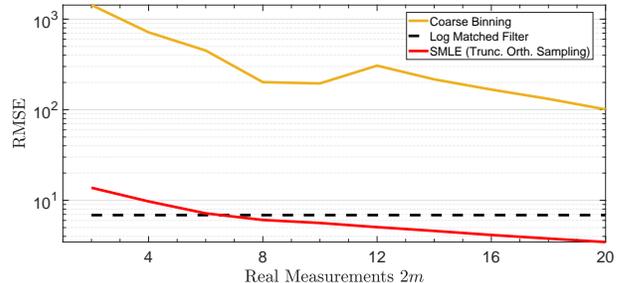}
\caption{The RMSE as a function of the number of real measurements ($2m$) for the polystyrene head dataset.}
\label{Fig: face RMSE}
\end{figure}

 Figure \ref{table: Head images} shows the reconstructed images of the sketched lidar, coarse binning and matched filter approaches, as well as the ground truth image. We first notice that our sketched lidar method sufficiently reconstructs the polystyrene head scene for all sketch sizes, even for the circular mean estimate ($m=1$) in (a). In contrast, the coarse binning approach fails for all corresponding measurements $\tilde{m}$ with significant staircase artifacts arising. Figure \ref{Fig: face RMSE} shows the RMSE, in comparison to the ground truth, as a function of the number of real measurements ($2m$). Here we omit the small proportion of pixels that consist of two peaks from the RMSE calculation for sake of fair comparison with the existing methods that can only estimate a single peak. We observe that our sketched lidar method produces a smaller RMSE as the measurement size increases and achieves a smaller RMSE than the LMF approach for larger measurements.  In comparison, the coarse binning method obtain estimates that produce a large RMSE consistently throughout. As such, this suggests that our sketched lidar approach does not compromise reduced resolution in favour of compression which is very apparent in the coarse binning method.

\begin{table}[ht!]
\centering
\begin{tabular}{c|c}
\textbf{\large Sketched Lidar} & \textbf{\large Coarse Binning}\\\\
\textbf{RMSE} $=10.63$ & \textbf{RMSE} $=1430.1$\\ 
 \includegraphics[scale=0.251]{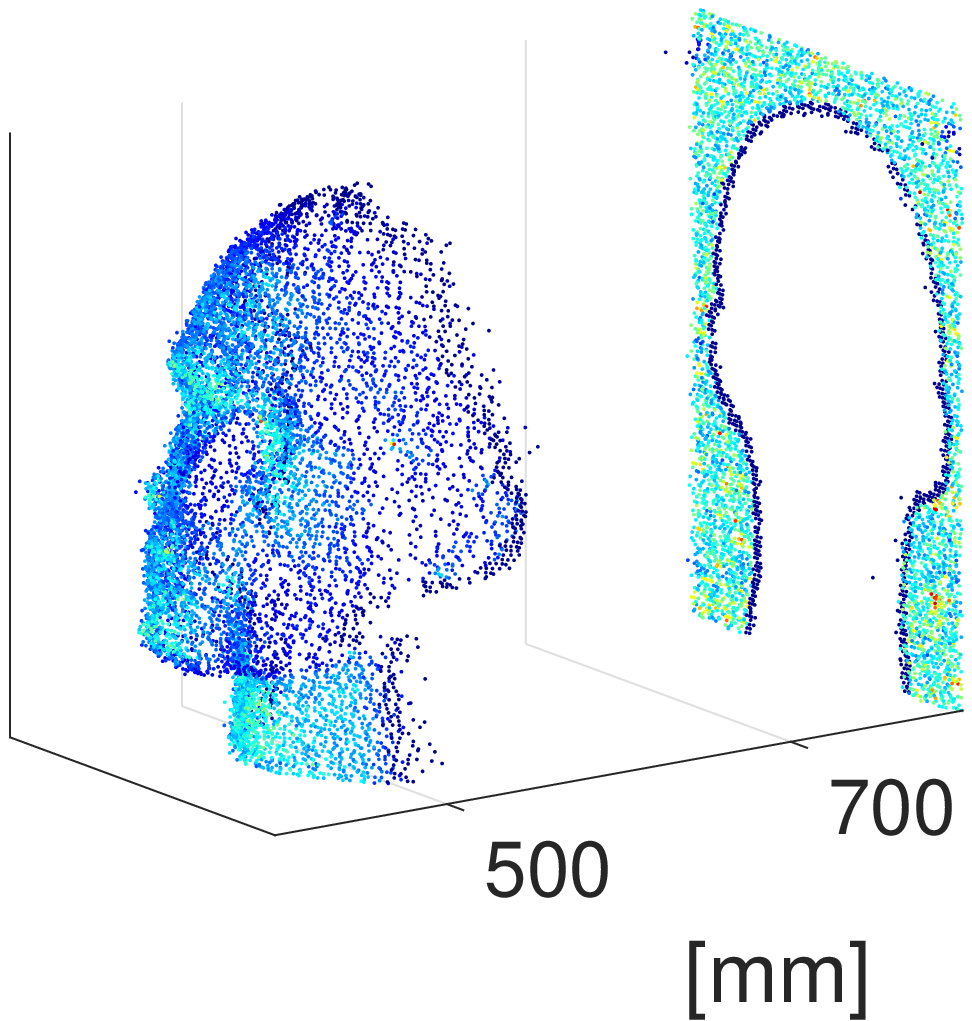}& \includegraphics[scale=0.2]{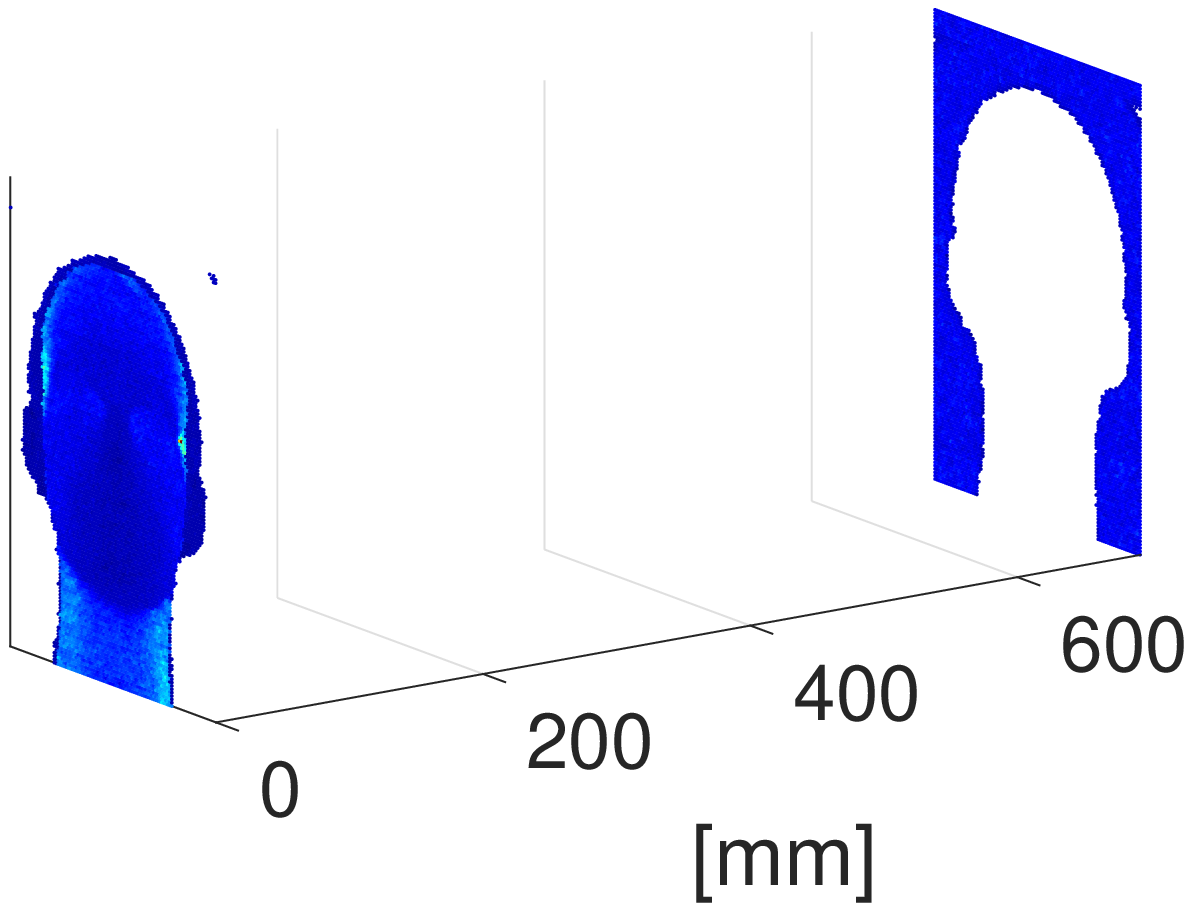} \\
Compression $= 0.0059$ & Compression $= 0.0059$\\ 
 a) Real Measurements $=2$ ($m=1$) &b) Measurements $=2$ \\\\
\textbf{RMSE} $=8.01$  & \textbf{RMSE} $=201.6$ \\ 
 \includegraphics[scale=0.251]{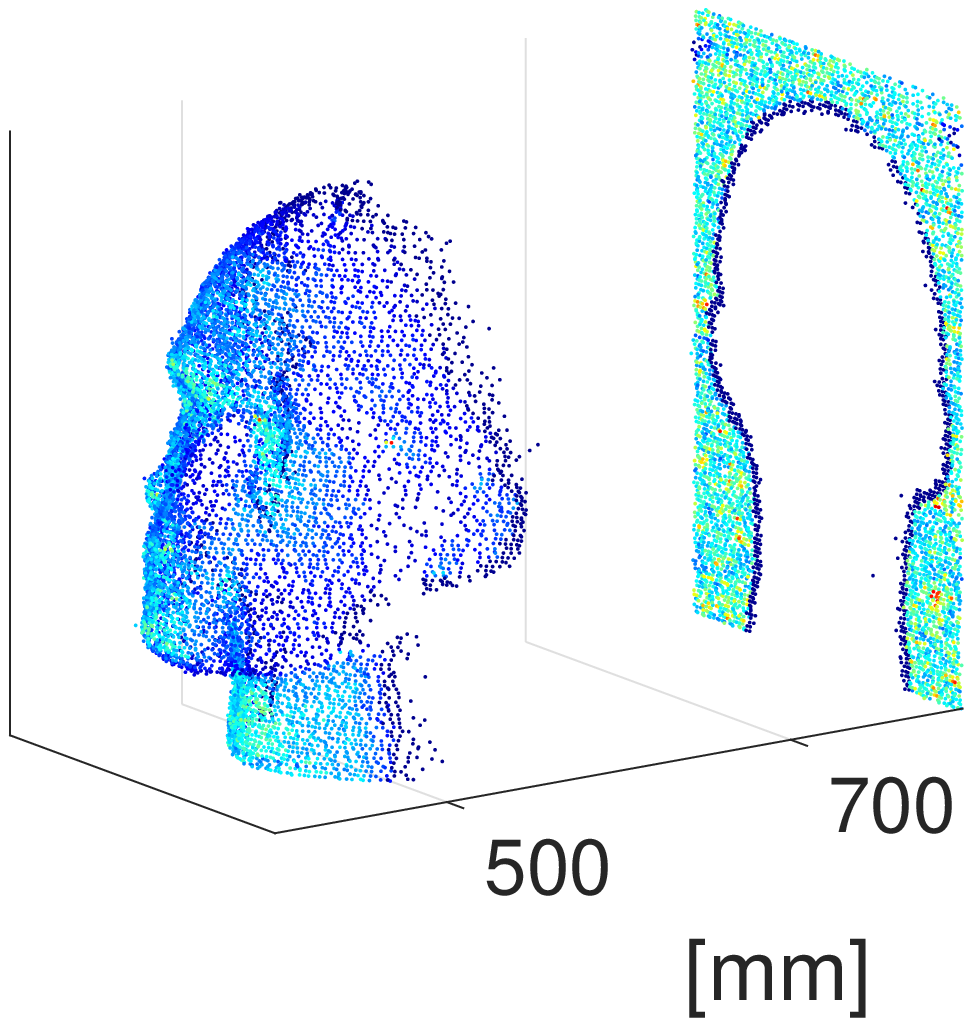}& \includegraphics[scale=0.2]{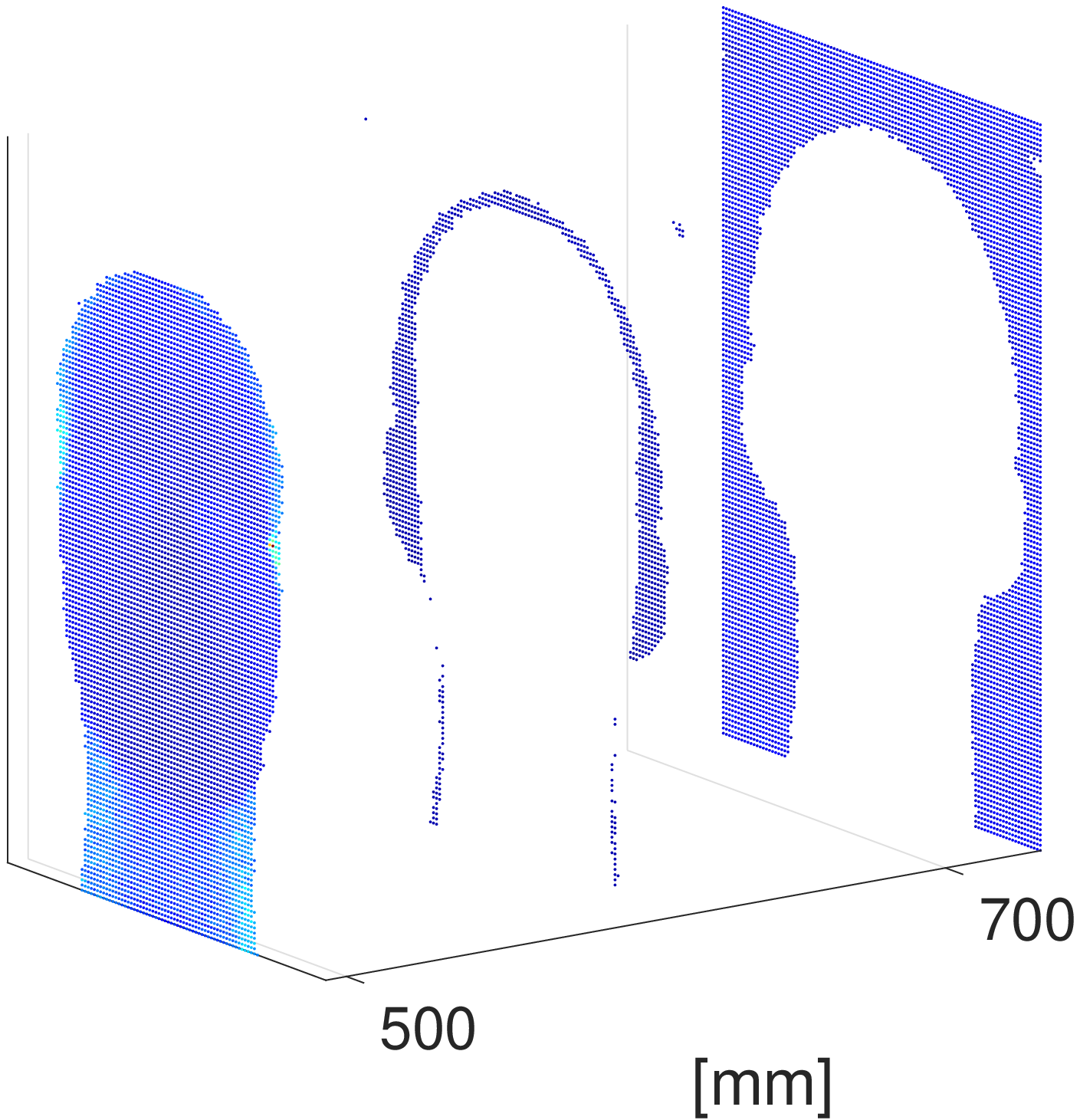} \\
Compression $= 0.0237$ & Compression $= 0.0237$\\
c) Real Measurements $=8$ ($m=4$) &d) Measurements $=8$ \\\\
\textbf{RMSE} $=7.40$  & \textbf{RMSE} $=100.9$ \\ 
 \includegraphics[scale=0.251]{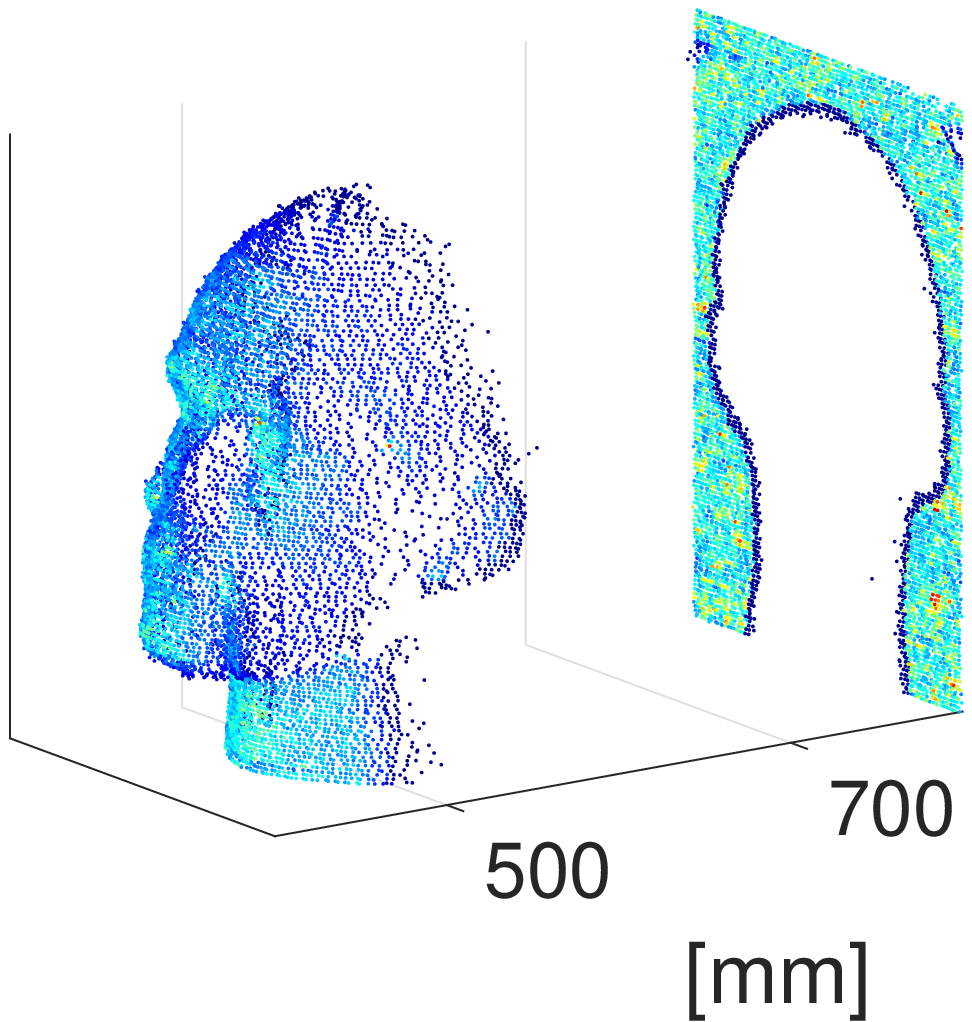}& \includegraphics[scale=0.2]{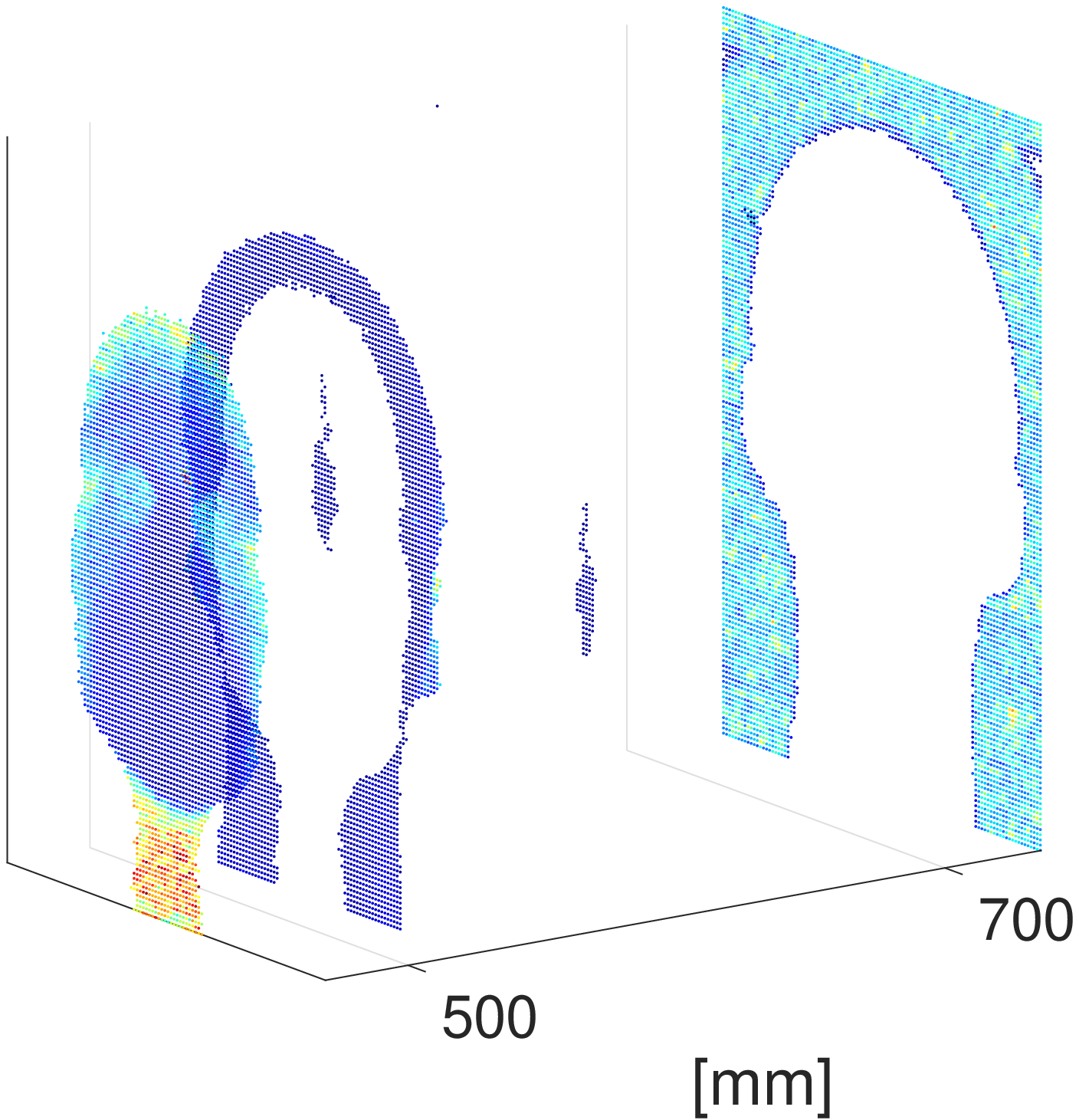} \\
Compression $= 0.0593$ & Compression $= 0.0593$\\
e) Real Measurements $=20$ ($m=10$) &f) Measurements $=20$ \\\\
\textbf{\large Matched Filter} & \textbf{\large Ground Truth}\\\\
\textbf{RMSE} $=6.87$  &  \\ 
 \includegraphics[scale=0.251]{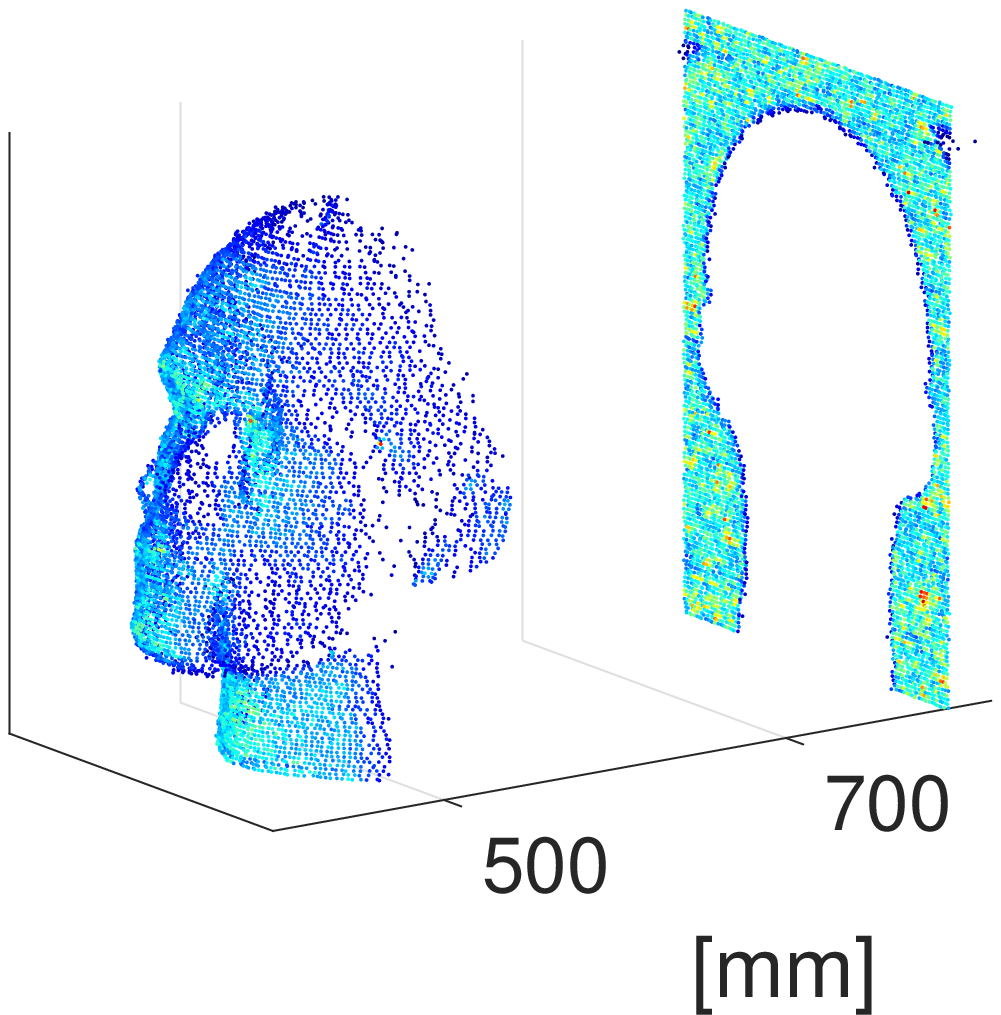}& \includegraphics[scale=0.251]{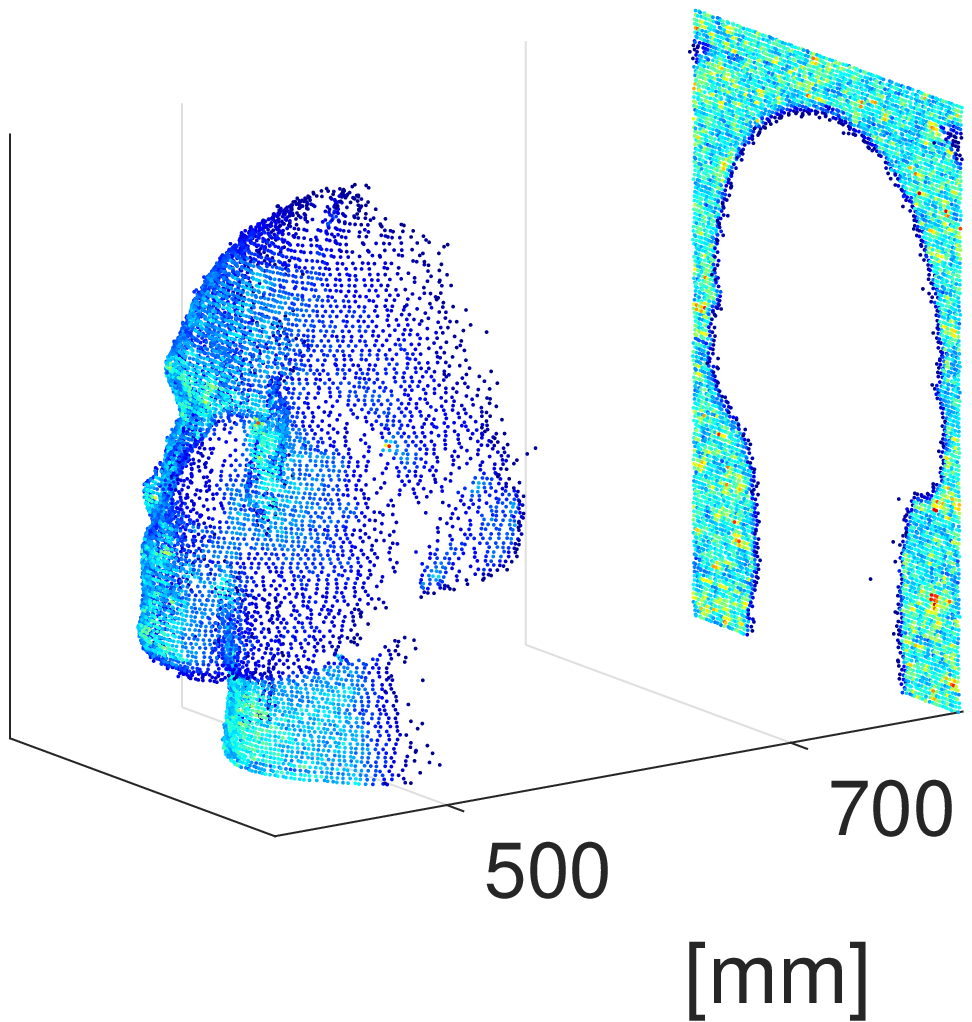} \\
 g) No Compression &h)\\
\multicolumn{2}{c}{\includegraphics[scale=0.4]{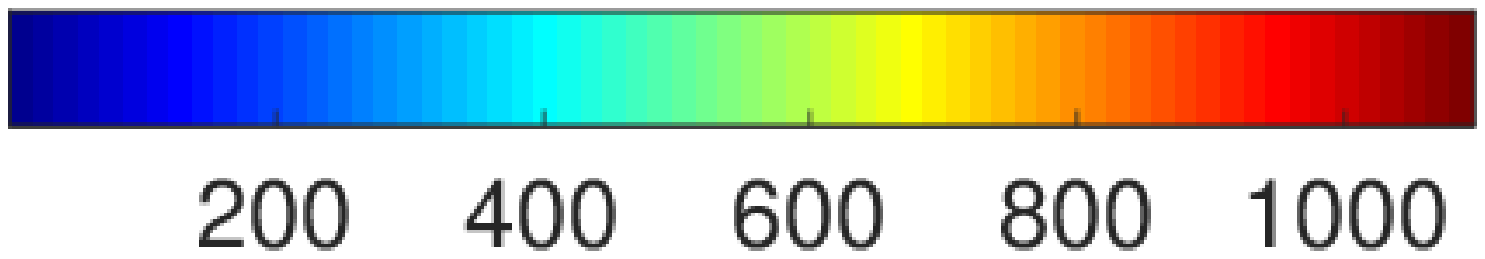}}\\ 
\multicolumn{2}{c}{  \qquad Intensity}\\
\end{tabular}
\captionof{figure}{The face dataset lidar reconstructions of the sketched lidar and coarse binning method for the real valued measurement size $2,8,20$. Both the matched filter reconstruction and the ground truth image are given for comparison. }
\label{table: Head images}
\end{table}

\subsubsection{Humans Behind Camouflage }\label{subsubsec: camo dataset}
\begin{figure}[ht!]
\centering
\includegraphics[width=0.48\textwidth]{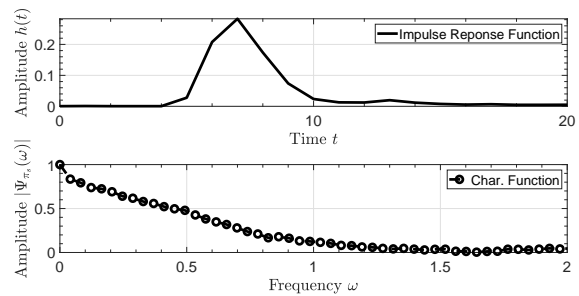}
\caption{ The CF (bottom) of the data driven impulse response function (top) of the camouflage dataset. }
\label{Fig: camo char func}
\end{figure}

The second scene consists of two humans standing behind a camouflage net approximately 320 metres away from the lidar device. Further details can be found of the scene in \cite{camo,tobin2017long}. The data cube has width and height of 32 pixels, $N_r=N_c=32$ and a total of $T=153$ time-stamps. A total acquisition time of 5.6 milliseconds was used for each pixel resulting in an average photon count of $\bar{n}=871$ with an approximate SBR of 2.35. The vast majority of pixels have 2 surfaces (the camouflage net and a human) where the net (first peak) accounts for the biggest intensity. The parameter set to be estimated for each pixel is $\theta=(t_1,t_2,\alpha_1,\alpha_2)$ of dimension 4. We compare our results with the full data EM algorithm as well as the coarse binning compression technique. For this experiment, the coarse binning algorithm uses the EM estimate once the data cube has been down-sampled as the matched filtering algorithm is only applicable to single peak cases. Due to the lack of a ground truth, we compare the reconstructions of the camouflage scene to the full data EM algorithm reconstruction and equate the relevant compression of both the sketched lidar framework and the coarse binning technique. The data driven impulse response function $h$ and its corresponding CF obtained from (\ref{Eqn: Char function Obs Model}), are shown in Figure \ref{Fig: camo char func}. Again, we only present the results for the truncated orthogonal sampling scheme, from Section \ref{subsec sampling schemes}, but we observed in practice that the alternative random orthogonal sampling scheme produces similar results. We uniformly sampled 10 starting points for each of peak $t_1$ and $t_2$ and initialised with the smallest sketched cost function from (\ref{Eqn: CL loss function}).

\par Figure \ref{table: Camo dataset} shows the reconstructed images of the sketched lidar, coarse binning and EM algorithm methods. Evidently, the reconstruction of our sketched lidar approach becomes better as the number of real measurements ($2m$) increases, for instance the torso of the human positioned near 600 cm has greater clarity in sketch size 20 compared to sketch size 4 where more spurious peaks are detected. However, the sketched lidar reconstruction for $m=2$ is still sufficient in comparison to the EM reconstruction in (g), while in contrast the coarse binning method fails to reconstruct either human for the corresponding number of measurements. The coarse binning method once again suffers from the stair case effect as seen by the lack of width of the first human standing at position 200 cm in (f). Furthermore, the compression due to the coarse binning results in poor depth accuracy as seen by the position of the camouflage net in reconstruction (b) which has a disparity of approximately 120 cm in comparison to the EM reconstruction. Once again, this suggests that our sketched lidar approach does not compromise reduced resolution in favour of compression which is apparent in the coarse binning method.

\begin{table}[ht!]
\centering
\begin{tabular}{c|c}
\textbf{\large Sketched Lidar} & \textbf{\large Coarse Binning}\\
 \includegraphics[scale=0.28]{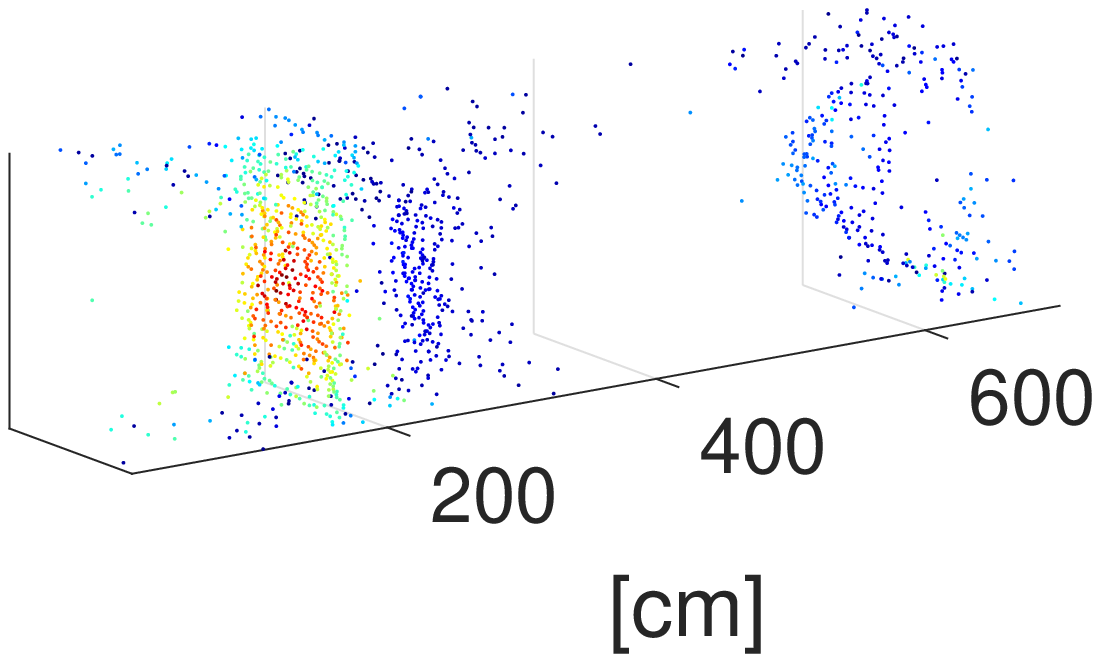}& \includegraphics[scale=0.26]{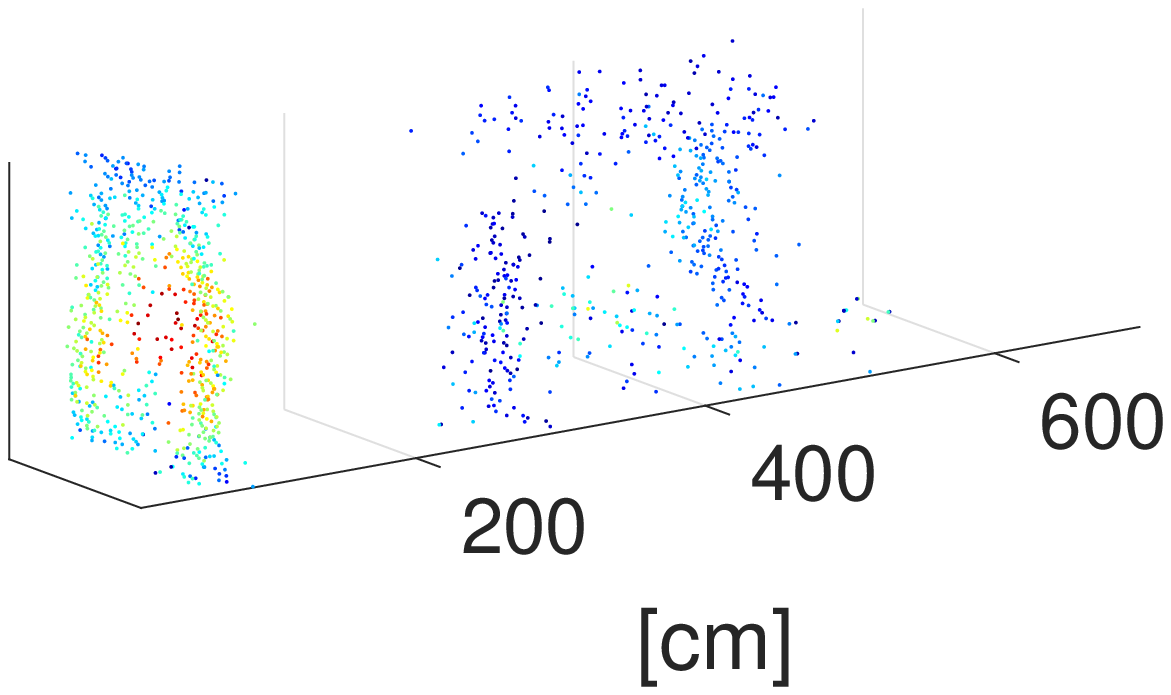} \\
Compression $= 0.0261$ & Compression $= 0.0261$\\ 
 a) Real Measurements $=4$ ($m=2$) &b) Measurements $=4$ \\
  \includegraphics[scale=0.28]{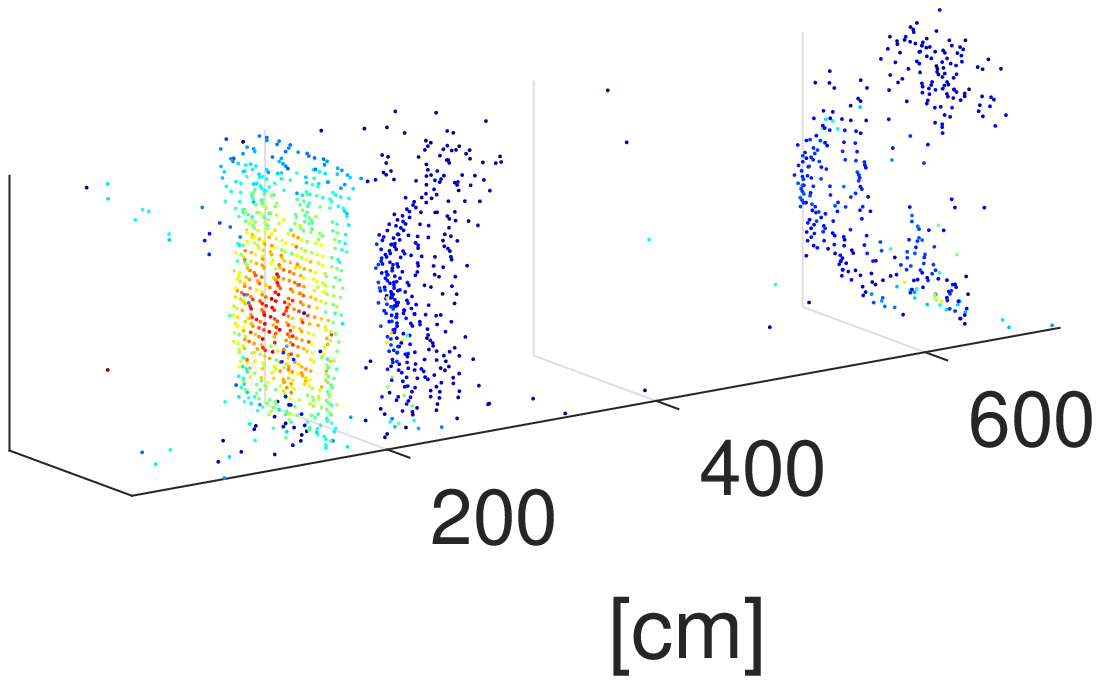}& \includegraphics[scale=0.26]{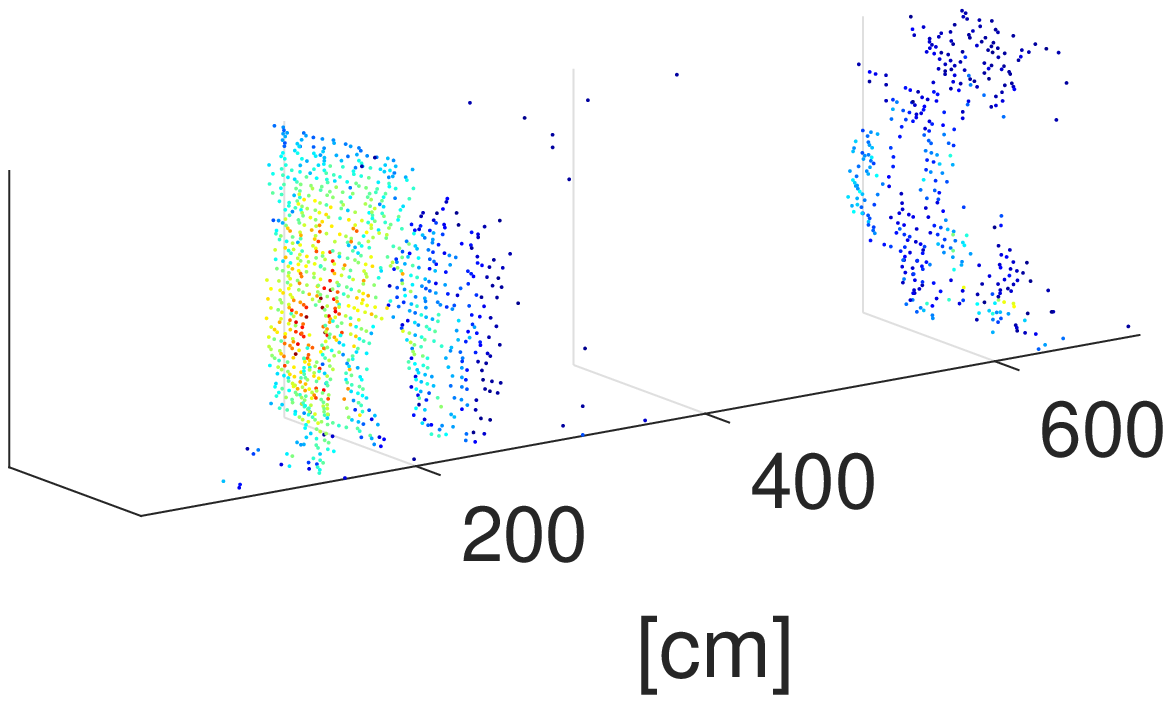} \\
Compression $= 0.0915$ & Compression $= 0.0915$\\ 
 c) Real Measurements $=14$ ($m=7$)  &d) Measurements $=14$\\
  \includegraphics[scale=0.26]{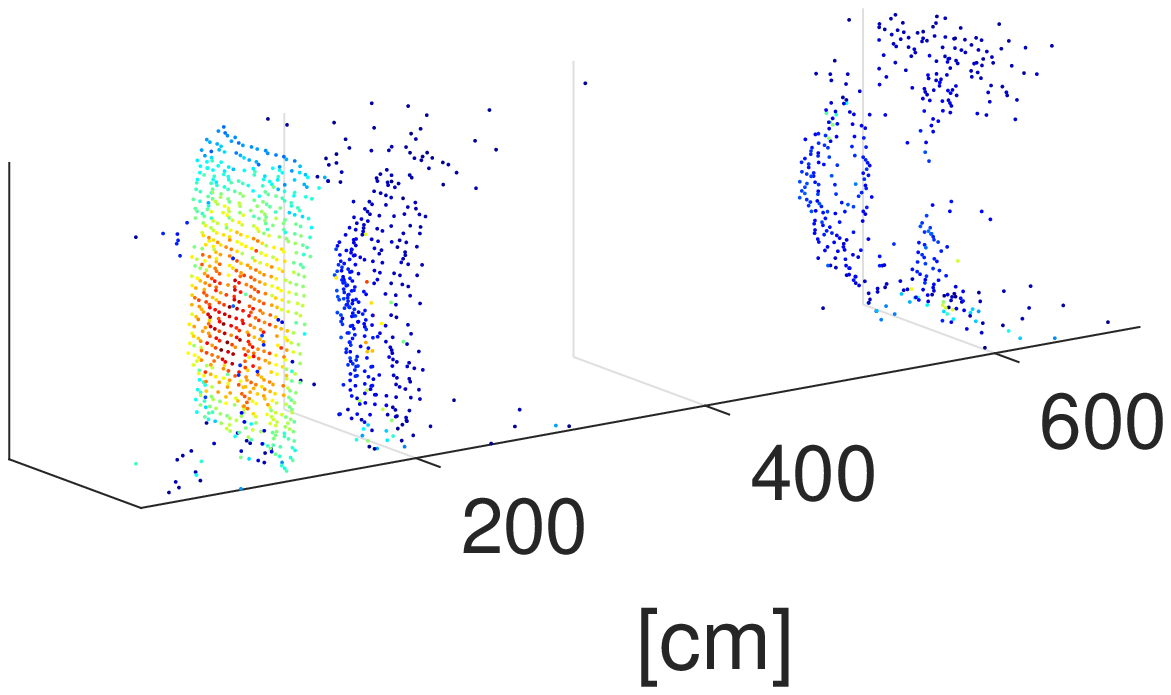}& \includegraphics[scale=0.26]{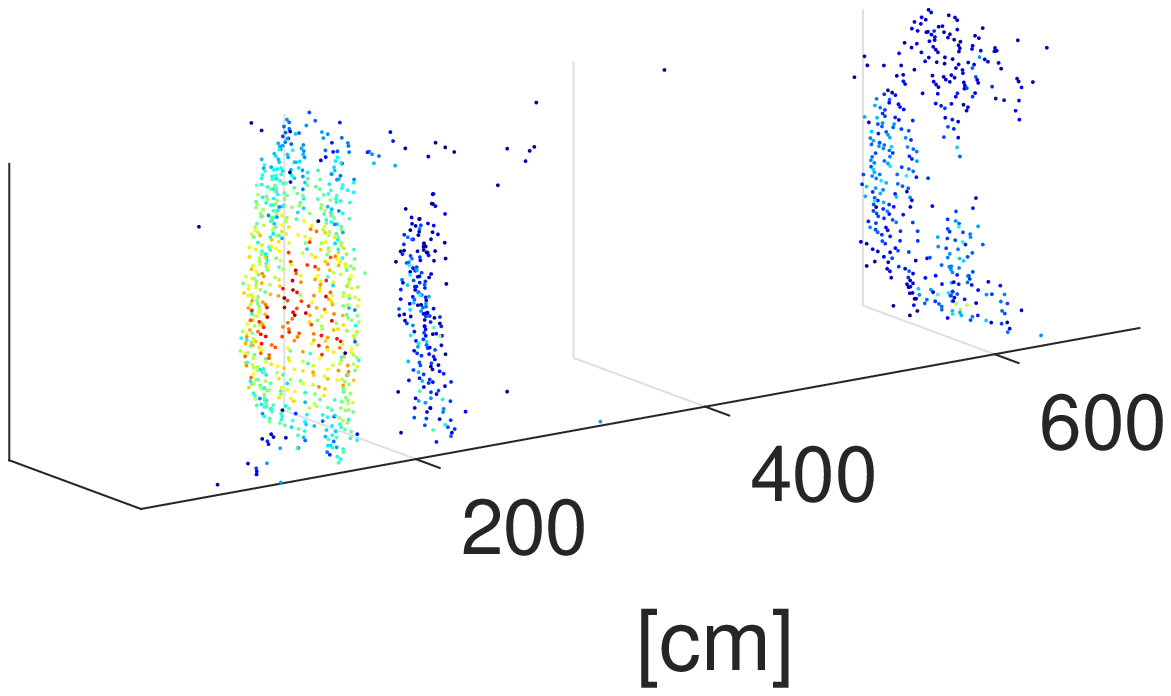} \\
 Compression $= 0.1307$ & Compression $= 0.1307$\\ 
 e) Real Measurements $=20$ ($m=10$) &f) Measurements $=20$\\
\noalign{\vskip 5mm} 
\multicolumn{2}{c}{\textbf{\large Expectation Maximization}}\\
\noalign{\vskip 5mm}
\multicolumn{2}{c}{\includegraphics[scale=0.3]{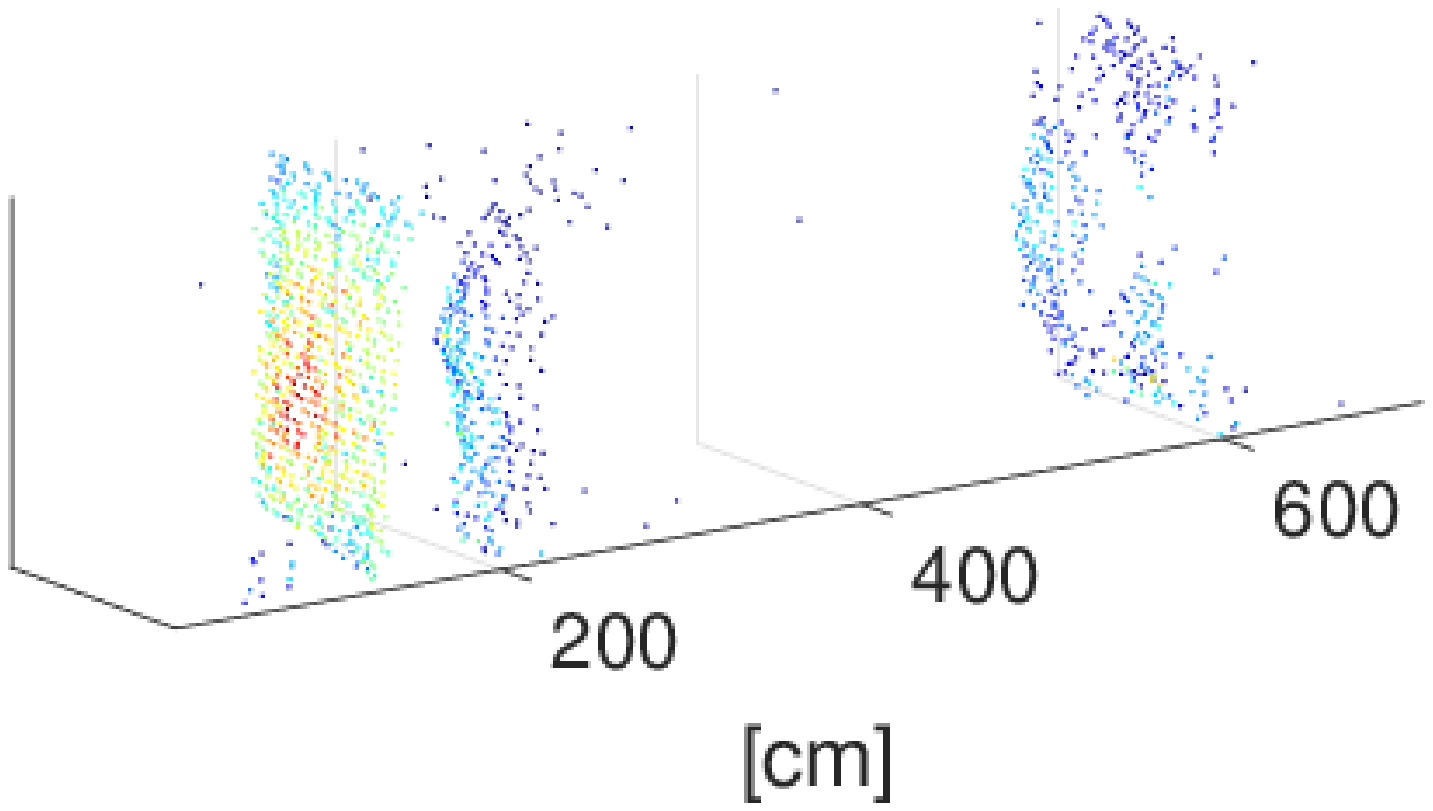}}\\
\noalign{\vskip 2mm}
\multicolumn{2}{c}{g) No Compression}\\
\multicolumn{2}{c}{\includegraphics[scale=0.35]{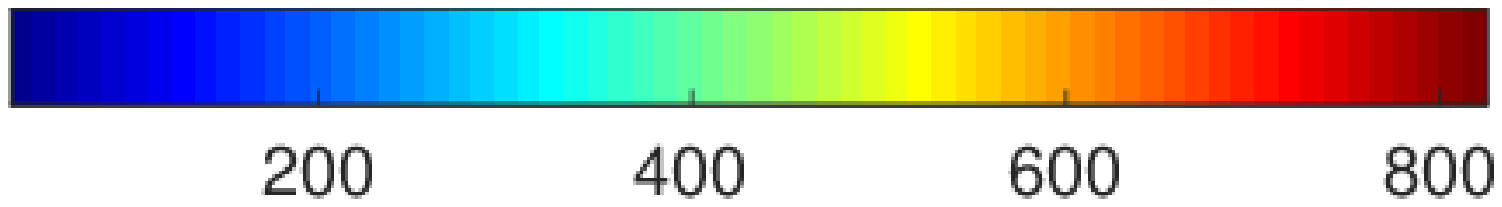}}\\ 
\multicolumn{2}{c}{ \qquad Intensity}\\
\end{tabular}
\captionof{figure}{The camouflage dataset lidar reconstructions of the sketched lidar and coarse binning method for the real valued measurement size ($2m$) of $2,8,20$. Both the matched filter reconstruction and the ground truth image are given for comparison.}
\label{table: Camo dataset}
\end{table}

\section{Conclusion}\label{Sec: Conclusion}
In this paper, we proposed a novel sketching solution to handle the major data processing bottleneck of single-photon lidar caused by the fine resolution of modern high rate, high resolution ToF image sensors. Our approach involved sampling the characteristic function of the observation model to form online statistics that have dimensionality proportional to the number of parameters of the model. Furthermore, we developed an efficient sketching algorithm, inspired by ECF estimation techniques, which has space and time complexity that fundamentally scales with the size of the sketch $m$, and is independent of both photon count and depth resolution. Two sampling schemes are proposed that sample in regions of the characteristic function that are \textit{blind} to photons originating from background sources. As a result, our method obtains estimates of the location and intensity parameters that are unbiased. Our novel sketch based acquisition removes the trade-off between depth resolution and data transfer complexity that is apparent in existing methods. Here we have only considered a simple pixel-wise depth estimate method in the form of the sketched MLE. It should be straightforward to incorporate the sketched statistics into more sophisticated state-of-the-art reconstruction algorithms, such as the real-time 3D algorithm in \cite{tachellaNComms} due to the Gaussian nature of the sketch statistics seen in Section \ref{subsec: CLT}. However, we leave this for future work. Another line of future work would be to use the sketch statistics for other algorithmic purposes such as target detection and multi-reflection detection.
\section*{Acknowledgements}
This work was supported by the ERC Advanced grant, project C-SENSE, (ERC-ADG-2015-694888). Mike. E. Davies is also supported by a Royal Society Wolfson Research Merit Award. The authors would like to thank the single-photon group at HWU
(\url{https://single-photon.com}) for the use of the datasets used in Section \ref{Subsec: Real Data}. The polystyrene head dataset in \ref{subsubsec: head dataset} and the camouflage dataset  in \ref{subsubsec: camo dataset} were obtained from \cite{manipop} and \cite{tachellaNComms}, respectively.

\section*{Code Availability}
 A MATLAB implementation of the SMLE algorithm is available at the repository \texttt{\url{https://gitlab.com/Tachella/sketched_lidar}}.

\bibliographystyle{IEEEtran}
\bibliography{main}

\begin{thebibliography}{10}
\providecommand{\url}[1]{#1}
\csname url@samestyle\endcsname
\providecommand{\newblock}{\relax}
\providecommand{\bibinfo}[2]{#2}
\providecommand{\BIBentrySTDinterwordspacing}{\spaceskip=0pt\relax}
\providecommand{\BIBentryALTinterwordstretchfactor}{4}
\providecommand{\BIBentryALTinterwordspacing}{\spaceskip=\fontdimen2\font plus
\BIBentryALTinterwordstretchfactor\fontdimen3\font minus
  \fontdimen4\font\relax}
\providecommand{\BIBforeignlanguage}[2]{{%
\expandafter\ifx\csname l@#1\endcsname\relax
\typeout{** WARNING: IEEEtran.bst: No hyphenation pattern has been}%
\typeout{** loaded for the language `#1'. Using the pattern for}%
\typeout{** the default language instead.}%
\else
\language=\csname l@#1\endcsname
\fi
#2}}
\providecommand{\BIBdecl}{\relax}
\BIBdecl

\bibitem{Hecht:18}
J.~Hecht, ``Lidar for self-driving cars,'' \emph{Opt. Photon. News}, vol.~29,
  no.~1, pp. 26--33, Jan 2018.

\bibitem{lidarauto}
J.~{Rapp}, J.~{Tachella}, Y.~{Altmann}, S.~{McLaughlin}, and V.~K. {Goyal},
  ``Advances in single-photon lidar for autonomous vehicles: Working
  principles, challenges, and recent advances,'' \emph{IEEE Signal Processing
  Magazine}, vol.~37, no.~4, pp. 62--71, 2020.

\bibitem{6159363}
J.~{Gao}, J.~{Sun}, J.~{Wei}, and Q.~{Wang}, ``Research of underwater target
  detection using a slit streak tube imaging lidar,'' in \emph{2011 Academic
  International Symposium on Optoelectronics and Microelectronics Technology},
  2011, pp. 240--243.

\bibitem{PIERZCHALA2018217}
M.~Pierzchała, P.~Giguère, and R.~Astrup, ``Mapping forests using an unmanned
  ground vehicle with 3{D} lidar and graph-slam,'' \emph{Computers and
  Electronics in Agriculture}, vol. 145, pp. 217 -- 225, 2018.

\bibitem{manipop}
J.~Tachella, Y.~Altmann, X.~Ren, A.~McCarthy, G.~S. Buller, S.~McLaughlin, and
  J.~Tourneret, ``Bayesian 3{D} reconstruction of complex scenes from
  single-photon lidar data,'' \emph{SIAM Journal on Imaging Sciences}, vol.~12,
  pp. 521--550, 03 2019.

\bibitem{Pawlikowska2017SinglephotonTI}
A.~M. Pawlikowska, A.~Halimi, R.~A. Lamb, and G.~S. Buller, ``Single-photon
  three-dimensional imaging at up to 10 kilometers range.'' \emph{Optics
  express}, vol. 25 10, pp. 11\,919--11\,931, 2017.

\bibitem{30frames}
C.~{Zhang}, S.~{Lindner}, I.~M. {Antolović}, J.~{Mata Pavia}, M.~{Wolf}, and
  E.~{Charbon}, ``A 30-frames/s, $252\times144$ {SPAD} flash lidar with 1728
  dual-clock 48.8-ps tdcs, and pixel-wise integrated histogramming,''
  \emph{IEEE Journal of Solid-State Circuits}, vol.~54, no.~4, pp. 1137--1151,
  2019.

\bibitem{3Dstacked}
S.~W. {Hutchings}, N.~{Johnston}, I.~{Gyongy}, T.~{Al Abbas}, N.~A.~W.
  {Dutton}, M.~{Tyler}, S.~{Chan}, J.~{Leach}, and R.~K. {Henderson}, ``A
  reconfigurable 3-{D}-stacked {SPAD} imager with in-pixel histogramming for
  flash lidar or high-speed time-of-flight imaging,'' \emph{IEEE Journal of
  Solid-State Circuits}, vol.~54, no.~11, pp. 2947--2956, 2019.

\bibitem{tachellaNComms}
J.~Tachella, Y.~Altmann, N.~Mellado, A.~McCarthy, R.~Tobin, G.~S. Buller,
  J.~Tourneret, and S.~McLaughlin, ``Real-time 3{D} reconstruction from
  single-photon lidar data using plug-and-play point cloud denoisers,''
  \emph{Nature communications}, vol.~10, no.~1, pp. 1--6, 2019.

\bibitem{hardware}
R.~K. {Henderson}, N.~{Johnston}, H.~{Chen}, D.~D. {Li}, G.~{Hungerford},
  R.~{Hirsch}, D.~{McLoskey}, P.~{Yip}, and D.~J.~S. {Birch}, ``A 192×128 time
  correlated single photon counting imager in 40nm {CMOS} technology,'' in
  \emph{ESSCIRC 2018 - IEEE 44th European Solid State Circuits Conference
  (ESSCIRC)}, 2018, pp. 54--57.

\bibitem{slidingGate}
X.~Ren, P.~W.~R. Connolly, A.~Halimi, Y.~Altmann, S.~McLaughlin, I.~Gyongy,
  R.~K. Henderson, and G.~S. Buller, ``High-resolution depth profiling using a
  range-gated {CMOS} {SPAD} quanta image sensor,'' \emph{Opt. Express},
  vol.~26, no.~5, pp. 5541--5557, Mar 2018.

\bibitem{128x96}
R.~J. {Walker}, J.~A. {Richardson}, and R.~K. {Henderson}, ``A 128×96 pixel
  event-driven phase-domain {$\Delta\Sigma$}-based fully digital 3{D} camera in
  0.13$\mu$m {CMOS} imaging technology,'' in \emph{2011 IEEE International
  Solid-State Circuits Conference}, 2011, pp. 410--412.

\bibitem{128x128}
F.~M. {Della Rocca}, H.~{Mai}, S.~W. {Hutchings}, T.~{Al Abbas}, A.~{Tsiamis},
  P.~{Lomax}, I.~{Gyongy}, N.~A.~W. {Dutton}, and R.~K. {Henderson}, ``A 128 ×
  128 {SPAD} dynamic vision-triggered time of flight imager,'' in \emph{ESSCIRC
  2019 - IEEE 45th European Solid State Circuits Conference (ESSCIRC)}, 2019,
  pp. 93--96.

\bibitem{128x128b}
F.~{Mattioli Della Rocca}, H.~{Mai}, S.~W. {Hutchings}, T.~A. {Abbas},
  K.~{Buckbee}, A.~{Tsiamis}, P.~{Lomax}, I.~{Gyongy}, N.~A.~W. {Dutton}, and
  R.~K. {Henderson}, ``A 128 × 128 {SPAD} motion-triggered time-of-flight
  image sensor with in-pixel histogram and column-parallel vision processor,''
  \emph{IEEE Journal of Solid-State Circuits}, vol.~55, no.~7, pp. 1762--1775,
  2020.

\bibitem{rapp2020dithered}
J.~Rapp, M.~A. Dawson, R., and V.~K. Goyal, ``Dithered depth imaging,''
  \emph{Optics Express}, vol.~28, no.~23, pp. 35\,143--35\,157, 2020.

\bibitem{7178153}
A.~{Kadambi} and P.~T. {Boufounos}, ``Coded aperture compressive 3-{D} lidar,''
  in \emph{2015 IEEE International Conference on Acoustics, Speech and Signal
  Processing (ICASSP)}, 2015, pp. 1166--1170.

\bibitem{halimi:hal-02298998}
A.~Halimi, P.~Ciuciu, A.~Mccarthy, S.~Mclaughlin, and G.~S. Buller, ``{Fast
  adaptive scene sampling for single-photon 3{D} lidar images},'' in
  \emph{{IEEE CAMSAP 2019 - International Workshop on Computational Advances in
  Multi-Sensor Adaptive Processing}}, Le Gosier (Guadeloupe), France, Dec.
  2019.

\bibitem{maksymova2018review}
I.~Maksymova, C.~Steger, and N.~Druml, ``Review of lidar sensor data
  acquisition and compression for automotive applications,'' in
  \emph{Multidisciplinary Digital Publishing Institute Proceedings}, vol.~2,
  no.~13, 2018, p. 852.

\bibitem{hansenGeMM}
L.~P. Hansen, ``Large sample properties of generalized method of moments
  estimators,'' \emph{Econometrica}, vol.~50, no.~4, pp. 1029--1054, 1982.

\bibitem{gemmhall}
A.~Hall, \emph{Generalized Method of Moments}, 11 2007, pp. 230 -- 255.

\bibitem{10.2307/2958763}
A.~Feuerverger and A.~Mureika, ``The empirical characteristic function and its
  applications,'' \emph{The Annals of Statistics}, vol.~5, no.~1, pp. 88--97,
  1977.

\bibitem{10.2307/2985144}
A.~Feuerverger and P.~McDunnough, ``On the efficiency of empirical
  characteristic function procedures,'' \emph{Journal of the Royal Statistical
  Society. Series B (Methodological)}, vol.~43, no.~1, pp. 20--27, 1981.

\bibitem{gribonval2020statistical}
R.~Gribonval, G.~Blanchard, N.~Keriven, and Y.~Traonmilin, ``Statistical
  learning guarantees for compressive clustering and compressive mixture
  modeling,'' \emph{arXiv preprint arXiv:2004.08085}, 2020.

\bibitem{keriven2018sketching}
N.~Keriven, A.~Bourrier, R.~Gribonval, and P.~P{\'e}rez, ``Sketching for
  large-scale learning of mixture models,'' \emph{Information and Inference: A
  Journal of the IMA}, vol.~7, no.~3, pp. 447--508, 2018.

\bibitem{SheehanCICA}
M.~P. {Sheehan}, M.~S. {Kotzagiannidis}, and M.~E. {Davies}, ``Compressive
  independent component analysis,'' in \emph{2019 27th European Signal
  Processing Conference (EUSIPCO)}, 2019, pp. 1--5.

\bibitem{Gyongy:20}
\BIBentryALTinterwordspacing
I.~Gyongy, S.~W. Hutchings, A.~Halimi, M.~Tyler, S.~Chan, F.~Zhu,
  S.~McLaughlin, R.~K. Henderson, and J.~Leach, ``High-speed 3{D} sensing via
  hybrid-mode imaging and guided upsampling,'' \emph{Optica}, vol.~7, no.~10,
  pp. 1253--1260, Oct 2020. [Online]. Available:
  \url{http://www.osapublishing.org/optica/abstract.cfm?URI=optica-7-10-1253}
\BIBentrySTDinterwordspacing

\bibitem{Krstajic:15}
\BIBentryALTinterwordspacing
N.~Krstaji\'{c}, S.~Poland, J.~Levitt, R.~Walker, A.~Erdogan, S.~Ameer-Beg, and
  R.~K. Henderson, ``0.5 billion events per second time correlated single
  photon counting using {C}{M}{O}{S} {SPAD} arrays,'' \emph{Opt. Lett.},
  vol.~40, no.~18, pp. 4305--4308, Sep 2015. [Online]. Available:
  \url{http://ol.osa.org/abstract.cfm?URI=ol-40-18-4305}
\BIBentrySTDinterwordspacing

\bibitem{Kirmani58}
\BIBentryALTinterwordspacing
A.~Kirmani, D.~Venkatraman, D.~Shin, A.~Cola{\c c}o, F.~N.~C. Wong, J.~H.
  Shapiro, and V.~K. Goyal, ``First-photon imaging,'' \emph{Science}, vol. 343,
  no. 6166, pp. 58--61, 2014. [Online]. Available:
  \url{https://science.sciencemag.org/content/343/6166/58}
\BIBentrySTDinterwordspacing

\bibitem{poissonmodel}
S.~{Hernandez-Marin}, A.~M. {Wallace}, and G.~J. {Gibson}, ``Bayesian analysis
  of lidar signals with multiple returns,'' \emph{IEEE Transactions on Pattern
  Analysis and Machine Intelligence}, vol.~29, no.~12, pp. 2170--2180, 2007.

\bibitem{altmann1}
Y.~{Altmann}, X.~{Ren}, A.~{McCarthy}, G.~S. {Buller}, and S.~{McLaughlin},
  ``Lidar waveform-based analysis of depth images constructed using sparse
  single-photon data,'' \emph{IEEE Transactions on Image Processing}, vol.~25,
  no.~5, pp. 1935--1946, 2016.

\bibitem{altmann2}
Y.~{Altmann} and S.~{McLaughlin}, ``Range estimation from single-photon lidar
  data using a stochastic {EM} approach,'' in \emph{2018 26th European Signal
  Processing Conference (EUSIPCO)}, 2018, pp. 1112--1116.

\bibitem{carrasco2000generalization}
M.~Carrasco and J.~P. Florens, ``Generalization of {GMM} to a continuum of
  moment conditions,'' \emph{Econometric Theory}, pp. 797--834, 2000.

\bibitem{osomeFouriermethods}
A.~Feuerverger and P.~McDunnough, ``On some {F}ourier methods for inference,''
  \emph{Journal of the American Statistical Association}, vol.~76, no. 374, pp.
  379--387, 1981.

\bibitem{lukacs1952analytic}
E.~Lukacs, O.~Sz{\'a}sz \emph{et~al.}, ``On analytic characteristic
  functions,'' \emph{Pacific J. Math}, vol.~2, no.~4, pp. 615--625, 1952.

\bibitem{Dirich_kernel}
A.~Bashirov, ``Chapter 12 - fourier series and integrals,'' in
  \emph{Mathematical Analysis Fundamentals}.\hskip 1em plus 0.5em minus
  0.4em\relax Boston: Elsevier, 2014, pp. 307--345.

\bibitem{jammalamadaka2001topics}
S.~R. Jammalamadaka and A.~Sengupta, \emph{Topics in circular
  statistics}.\hskip 1em plus 0.5em minus 0.4em\relax world scientific, 2001,
  vol.~5.

\bibitem{detectSBR}
\BIBentryALTinterwordspacing
P.~Padmanabhan, C.~Zhang, and E.~Charbon, ``Modeling and analysis of a direct
  time-of-flight sensor architecture for {L}i{DAR} applications,''
  \emph{Sensors}, vol.~19, no.~24, 2019. [Online]. Available:
  \url{https://www.mdpi.com/1424-8220/19/24/5464}
\BIBentrySTDinterwordspacing

\bibitem{eldar_kutyniok_2012}
Y.~C. Eldar and G.~Kutyniok, \emph{Compressed sensing: theory and
  applications}.\hskip 1em plus 0.5em minus 0.4em\relax Cambridge university
  press, 2012.

\bibitem{keriven_conference}
N.~{Keriven}, A.~{Bourrier}, R.~{Gribonval}, and P.~{Pérez}, ``Sketching for
  large-scale learning of mixture models,'' in \emph{2016 IEEE International
  Conference on Acoustics, Speech and Signal Processing (ICASSP)}, 2016, pp.
  6190--6194.

\bibitem{cordic_ref}
\BIBentryALTinterwordspacing
F.~de~Dinechin, M.~Istoan, and G.~Sergent, ``Fixed-point trigonometric
  functions on {FPGA}s,'' \emph{{SIGARCH} Comput. Archit. News}, vol.~41,
  no.~5, p. 83–88, Jun. 2014. [Online]. Available:
  \url{https://doi.org/10.1145/2641361.2641375}
\BIBentrySTDinterwordspacing

\bibitem{schellekens2021asymmetric}
V.~Schellekens and L.~Jacques, ``Asymmetric compressive learning guarantees
  with applications to quantized sketches,'' \emph{arXiv preprint
  arXiv:2104.10061}, 2021.

\bibitem{10.2307/1912775}
L.~Hansen, ``Large sample properties of generalized method of moments
  estimators,'' \emph{Econometrica}, vol.~50, no.~4, pp. 1029--1054, 1982.

\bibitem{hansen2step}
L.~P. Hansen, J.~Heaton, and A.~Yaron, ``Finite-sample properties of some
  alternative {GMM} estimators,'' \emph{Journal of Business $\&$ Economic
  Statistics}, vol.~14, no.~3, pp. 262--280, 1996.

\bibitem{HAUSMAN201145}
J.~Hausman, R.~Lewis, K.~Menzel, and W.~Newey, ``Properties of the {CUE}
  estimator and a modification with moments,'' \emph{Journal of Econometrics},
  vol. 165, no.~1, pp. 45 -- 57, 2011, moment Restriction-Based Econometric
  Methods.

\bibitem{VivekLidar}
D.~{Shin}, A.~{Kirmani}, V.~K. {Goyal}, and J.~H. {Shapiro}, ``Photon-efficient
  computational 3-{D} and reflectivity imaging with single-photon detectors,''
  \emph{IEEE Transactions on Computational Imaging}, vol.~1, no.~2, pp.
  112--125, 2015.

\bibitem{RappPhotonEfficient}
J.~Rapp and V.~Goyal, ``A few photons among many: Unmixing signal and noise for
  photon-efficient active imaging,'' \emph{IEEE Transactions on Computational
  Imaging}, vol.~PP, 09 2016.

\bibitem{FisherEfficiency}
R.~A. Fisher, ``On the mathematical foundations of theoretical statistics,''
  \emph{Philosophical Transactions of the Royal Society of London. Series A,
  Containing Papers of a Mathematical or Physical Character}, vol. 222, pp.
  309--368, 1922.

\bibitem{amari2007methods}
H.~Amari, S.and~Nagaoka, \emph{Methods of information geometry}.\hskip 1em plus
  0.5em minus 0.4em\relax American Mathematical Soc., 2007, vol. 191.

\bibitem{logmatchedfilter}
G.~{Turin}, ``An introduction to matched filters,'' \emph{IRE Transactions on
  Information Theory}, vol.~6, no.~3, pp. 311--329, 1960.

\bibitem{EMAlg}
A.~P. Dempster, N.~M. Laird, and D.~B. Rubin, ``Maximum likelihood from
  incomplete data via the {EM} algorithm,'' \emph{Journal of the Royal
  Statistical Society. Series B (Methodological)}, vol.~39, no.~1, pp. 1--38,
  1977.

\bibitem{LidarFoliage}
R.~M. Marino and W.~R. Davis, ``Jigsaw: a foliage-penetrating 3{D} imaging
  laser radar system,'' \emph{Lincoln Lab J.}, vol.~15, no.~1, pp. 23--36,
  2005.

\bibitem{camo}
A.~{Halimi}, R.~{Tobin}, A.~{McCarthy}, S.~{McLaughlin}, and G.~S. {Buller},
  ``Restoration of multilayered single-photon 3{D} lidar images,'' in
  \emph{2017 25th European Signal Processing Conference (EUSIPCO)}, 2017, pp.
  708--712.

\bibitem{sheehan2021surface}
M.~P. Sheehan, J.~Tachella, and M.~E. Davies, ``Surface detection for sketched
  single photon lidar,'' in \emph{2021 29th European Signal Processing
  Conference ({EUSIPCO})}.\hskip 1em plus 0.5em minus 0.4em\relax IEEE, 2021,
  pp. 1--5.

\bibitem{tobin2017long}
R.~Tobin, A.~Halimi, A.~McCarthy, X.~Ren, K.~J. McEwan, S.~McLaughlin, and
  G.~S. Buller, ``Long-range depth profiling of camouflaged targets using
  single-photon detection,'' \emph{Optical Engineering}, vol.~57, no.~3, p.
  031303, 2017.

\end{thebibliography}

\appendices
\section{Deriving the Circular Mean Estimate From the ECF Estimation}\label{Appendix: Circ mean}
Given a single frequency $\omega\in\R$, we can define the sketch as $z_n=\frac{1}{n}\sum^n_{j=1}e^{{\rm i}\omega x_j}$ and the goal is to solve:
\begin{equation}
\label{Eqn: motivating example}
    \hat{\theta}=\argmin_\theta (z_n-\Psi_\pi(\omega))^2.
\end{equation}
Clearly, (\ref{Eqn: motivating example}) is minimised when $\Psi_\pi(\omega)=z_n$ and equating the real and complex components we get:

\begin{align}
  & \alpha e^{\frac{(\omega t)^2}{2}}\cos(\omega t)-(1-\alpha)D_{\frac{T-1}{2}}(\omega) = \frac{1}{n}\sum^n_{j=1}\cos(\omega x_j) \\
      & \alpha e^{\frac{(\omega t)^2}{2}}\sin(\omega t) = \frac{1}{n}\sum^n_{j=1}\sin(\omega x_j).
\end{align}
\par Notably, we can optimally choose the frequency to be $\omega=\frac{2\pi}{T}$ resulting in $D_{\frac{T-1}{2}}(\omega)=0$ and thereby ensure the characteristic function is sampled in a region where the background noise is not present. Consequently, dividing (28) by (29) we get 
\begin{equation}
  \dfrac{\alpha e^{\frac{(\frac{2\pi t}{T})^2}{2}}\cos(\frac{2\pi t}{T})}{\alpha e^{\frac{(\frac{2\pi t}{T})^2}{2}}\sin(\frac{2\pi t}{T})} = \dfrac{\sum^n_{j=1}\cos(\omega x_j)}{\sum^n_{j=1}\sin(\omega x_j)},  
\end{equation}
resulting in an optimal estimate of
\begin{equation}
    \label{Eqn: optimal cylic mean}
    \theta^*= \frac{T}{2 \pi} \phase \left\{ \sum_{j=1}^n \cos \left(\frac{2\pi x_j}{T}\right) + {\rm i}\sum_{j=1}^n \sin \left(\frac{2\pi x_j}{T} \right)\right\}
\end{equation}

\section{Photon Starved Regime}\label{App: Photon Starved Regime}
We evaluate the performance of our proposed sketched lidar method in the photon starved regime in comparison to transferring the photon time-stamps directly off-chip and estimating the depth of the surface. For fair comparison, we let $2m=n$ for each photon count $n$ in the photon starved regime. Both the matched filter and maximum peak estimate the depth location using the full photon count. Here we simulate a pixel of a lidar scene with a time window of $T=100$ using a Gaussian IRF with pulse width $\sigma=0.03T$ for photon counts $n=[1,3,5\dots,15]$ and SBR varying between 0.01 and 100. For each photon count and SBR pair, 1000 Monte-Carlo simulations were executed with randomly chosen depth position $t_0\in[1,2,\dots,T]$ and the RMSE was calculated. 

\begin{figure}[ht!]
\centering
\includegraphics[width=1\linewidth]{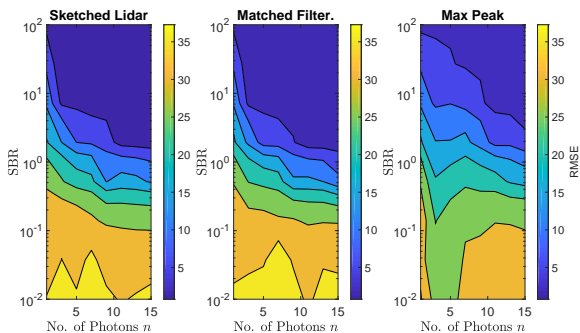}
 \captionsetup{type = figure}
 \caption{Sketched Lidar performs comparatively well (in terms of RMSE) compared with the full data approaches of matched filtering and maximum peak estimation in the photon starved regime.}
\label{fig: appendix starved photon regime 1}
\end{figure}

\begin{figure}[ht!]
\centering
\includegraphics[width=0.97\linewidth]{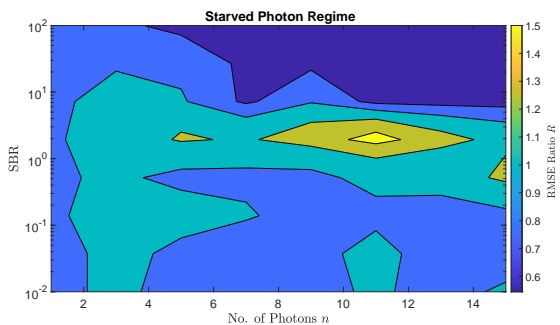}
 \captionsetup{type = figure}
 \caption{Comparison of the depth reconstruction of sketched lidar and matched filter using the RMSE ratio $R$ for varying SBR levels and photon counts in the photon starved regime. Sketched Lidar performs favourably compared to matched filtering for the majority of SBR values.}
\label{fig: appendix starved photon regime 2}
\end{figure}

\par Furthermore, we use the RMSE ratio between the sketched lidar and matched filter depth estimation, defined as 
 \begin{equation}
     R=\frac{\text{RMSE}_{\text{sketch}}}{\text{RMSE}_{\text{MF}}},
 \end{equation}
 where $\text{RMSE}_{\text{sketch}}$ and $\text{RMSE}_{\text{MF}}$ denote the RMSE of the sketched lidar and matched filter estimation, respectively. An $R>1$, indicates that the matched filter achieves on average a smaller RMSE than sketched lidar. Similarly an $R<1$,  indicates the sketched lidar estimation achieves on average a smaller RMSE than matched filter approach. Figures \ref{fig: appendix starved photon regime 1} and \ref{fig: appendix starved photon regime 2} show that the proposed sketched lidar approach does not suffer from a drop in estimation performance in both the photon starved regime and in the case of extremely low SBR in comparison with the matched filter that estimates the depth using all the detected photons.

\section{Comparison to the iFFT approach}\label{appendix: comparison to transient imag}
\begin{figure}[ht!]
\centering
\includegraphics[width=0.97\linewidth]{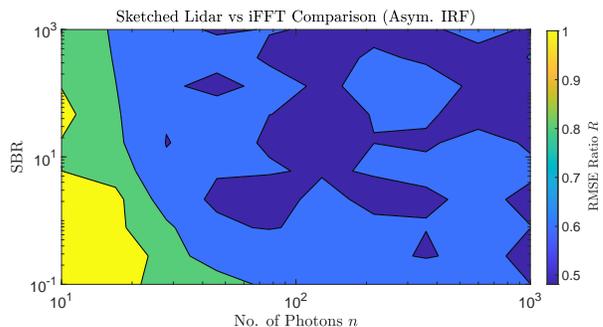}
 \captionsetup{type = figure}
 \caption{Comparison of the depth reconstruction of sketched lidar and the iFFT method using the RMSE ratio $R$ for varying SBR levels and photon counts for a non-Gaussian asymmetric IRF. Sketched Lidar performs equally or favourably to iFFT for all SBR and photon count pairs.}
\label{fig: appendix trans imaging fig}
\end{figure}

In Section \ref{Subsec: Syn Data}, we compared our proposed sketched lidar approach to the iFFT approach. The iFFT approach cannot incorporate information about the impulse response function while in the sketched lidar method the impulse response function is integrated throughout. To demonstrate this, we compare the performance of the sketched lidar and iFFT techniques for the non-Gaussian asymmetric IRF used in Section \ref{subsubsec: head dataset} (See Figure \ref{Fig: face char func}). For a signal-to-background ratio varying between 0.1-100 and a photon count ranging between 10-1000, a pixel from a lidar scene was simulated with randomly chosen depth position between $1,\dots,T$. A total of 1000 Monte-Carlo experiments were simulated for each SBR/photon count value with the RMSE recorded. For fair comparison we include an asymmetric correction for the iFFT approach to offset the bias of the asymmetric impulse response function. In Figure \ref{fig: appendix trans imaging fig}, the ratio between the RMSE of the sketch results and the RMSE of the iFFT estimation, for e.g.:
\begin{equation}
   R =  \frac{\text{RMSE}_{\text{sketch}}}{\text{RMSE}_{\text{iFFT}}}
\end{equation}
is displayed for $m=2$.

\par The improvement using the sketched lidar method over the iFFT approach is apparent. For the majority of the SBR/photon count pairs the sketched lidar method achieves approximately half the RMSE of that of the iFFT approach, highlighting the lack of information of the IRF the iFFT approach has incorporated into its depth estimation. 

\end{document}